\documentclass{emulateapj}

\usepackage{times}

\newcommand{\sfr}{\hbox{$\psi$}} 
\newcommand{\ssfr}{\hbox{$\psi_{\mathrm S}$}}
\newcommand{\tauv}{\hbox{$\hat{\tau}_{V}$}}

\shorttitle{Star formation histories of quiescent galaxies}
\shortauthors{Pacifici et al.}

\begin{document}

\title{The evolution of star formation histories of quiescent galaxies}

\author{Camilla~Pacifici,\altaffilmark{1,$\dagger$} Susan~A.~Kassin,\altaffilmark{2} Benjamin J. Weiner,\altaffilmark{3} Bradford Holden,\altaffilmark{4} Jonathan P. Gardner,\altaffilmark{1} Sandra M. Faber,\altaffilmark{4} Henry C. Ferguson,\altaffilmark{2} David C. Koo,\altaffilmark{4} Joel R. Primack,\altaffilmark{4} Eric F. Bell,\altaffilmark{5} Avishai Dekel,\altaffilmark{6} Eric Gawiser,\altaffilmark{7} Mauro Giavalisco,\altaffilmark{8} Marc Rafelski,\altaffilmark{2} Raymond C. Simons,\altaffilmark{9} Guillermo Barro,\altaffilmark{10} Darren J. Croton,\altaffilmark{11} Romeel Dav\'e,\altaffilmark{12,13,14} Adriano Fontana,\altaffilmark{15} Norman A. Grogin,\altaffilmark{2} Anton M. Koekemoer,\altaffilmark{2} Seong-Kook Lee,\altaffilmark{16} Brett Salmon,\altaffilmark{2} Rachel Somerville,\altaffilmark{7} Peter Behroozi\altaffilmark{10,$\ddagger$}}

\altaffiltext{1}{Goddard Space Flight Center, Code 665, Greenbelt, MD, USA}
\altaffiltext{2}{Space Telescope Science Institute, Baltimore, MD 21218, USA}
\altaffiltext{3}{Steward Observatory, 933 N. Cherry Ave., University of Arizona, Tucson, AZ 85721, USA}
\altaffiltext{4}{UCO/Lick Observatory, Department of Astronomy and Astrophysics, University of California, Santa Cruz, CA, USA}
\altaffiltext{5}{Department of Astronomy, University of Michigan, 500 Church St., Ann Arbor, MI 48109, USA}
\altaffiltext{6}{Center for Astrophysics and Planetary Science, Racah Institute of Physics, The Hebrew University, Jerusalem 91904, Israel}
\altaffiltext{7}{Department of Physics and Astronomy, Rutgers University, Piscataway, NJ 08854, USA}
\altaffiltext{8}{Astronomy Department, University of Massachusetts, Amherst, MA 01003, USA}
\altaffiltext{9}{Department of Physics and Astronomy, The Johns Hopkins University, 366 Bloomberg Center, Baltimore, MD 21218}
\altaffiltext{10}{Department of Astronomy, University of California, Berkeley, CA, USA}
\altaffiltext{11}{Centre for Astrophysics \& Supercomputing, Swinburne University of Technology, PO Box 218, Hawthorn, Victoria, 3122, Australia}
\altaffiltext{12}{University of the Western Cape, Bellville, Cape Town 7535, South Africa}
\altaffiltext{13}{South African Astronomical Observatories, Observatory, Cape Town 7525, South Africa}
\altaffiltext{14}{African Institute for Mathematical Sciences, Muizenberg, Cape Town 7545, South Africa}
\altaffiltext{15}{INAF Osservatorio  Astronomico  di  Roma,  via  Frascati  33, 00040 Monteporzio, Italy}
\altaffiltext{16}{Center for the Exploration of the Origin of the Universe, Department of Physics and Astronomy, Seoul National University, Seoul, 151-742, Republic of Korea}
\altaffiltext{$\dagger$}{NASA Postdoctoral Program Fellow}
\altaffiltext{$\ddagger$}{Hubble Fellow}

\begin{abstract}

Although there has been much progress in understanding how galaxies evolve, we still do not understand how and when they stop forming stars and become quiescent. We address this by applying our galaxy spectral energy distribution models, which incorporate physically motivated star formation histories (SFHs) from cosmological simulations, to a sample of quiescent galaxies at $0.2<z<2.1$. A total of 845 quiescent galaxies with multi-band photometry spanning rest-frame ultraviolet through near-infrared wavelengths are selected from the CANDELS dataset.
We compute median SFHs of these galaxies in bins of stellar mass and redshift.
At all redshifts and stellar masses, the median SFHs rise, reach a peak, and then decline to reach quiescence.
At high redshift, we find that the rise and decline are fast, as expected because the Universe is young.
At low redshift, the duration of these phases depends strongly on stellar mass. Low-mass galaxies ($\log(M_{\ast}/M_{\odot})\sim9.5$) grow on average slowly, take a long time to reach their peak of star formation ($\gtrsim 4$ Gyr), and the declining phase is fast ($\lesssim 2$ Gyr).
Conversely, high-mass galaxies ($\log(M_{\ast}/M_{\odot})\sim11$) grow on average fast ($\lesssim 2$ Gyr), and, after reaching their peak, decrease the star formation slowly ($\gtrsim 3$ Gyr).
These findings are consistent with galaxy stellar mass being a driving factor in determining how evolved galaxies are, with high-mass galaxies being the most evolved at any time (i.e., downsizing).
The different durations we observe in the declining phases also suggest that low- and high-mass galaxies experience different quenching mechanisms that operate on different timescales.

\end{abstract}


\section{Introduction}
\label{sec:intro}

Over the past decade, multi-wavelength observational surveys and cosmological simulations of galaxies have greatly improved our understanding of the formation and evolution of galaxies. Observations of colors, morphology, spectral type, and star formation of galaxies show a clear bimodality in the population, where \textit{blue} star-forming galaxies are separated from \textit{red} quiescent galaxies. One key unknown is how and when galaxies stop forming stars and become ``red and dead''.

The classical picture for red, early-type galaxies is that they form in a single-burst very early in the Universe (\citealt{baade1963}). This theory was successfully challenged. By looking at spectral indices of galaxies, \citet{worthey1992}, \citet{faber1995}, and \citet{trager2000} found that early-type galaxies span a large range of ages and thus showed that blue galaxies can become red via different mechanisms occurring at different times and masses. Since then, other parameters have been studied to explain the variety of histories observed. \citet{graves2009a} (see also \citealt{graves2009b,graves2010a,graves2010b}) explored the ages and metallicities of red galaxies and found correlations with not only stellar mass, but also radius, velocity dispersion, and surface brightness. Furthermore, \citet{cheung2012}, \citet{fang2013}, and \citet{barro2015} observed that the path to quiescence is strongly correlated with central surface density. 

How and when star-forming galaxies move to the red sequence is however still a matter of debate. A combination of different physical processes can explain the evolution of luminosity functions and color-magnitude diagrams (\citealt{willmer2006}, \citealt{bell2006}, \citealt{faber2007}, \citealt{skelton2012}). Passive evolution of the $z\sim 1$ quiescent population has been disfavored to explain the age-mass and metallicity-mass relations of local galaxies by \citet{harker2006}, \citet{schiavon2006}, \citet{ruhland2009}, and \citet{gallazzi2014}. However, \citet{choi2014} and \citet{shetty2015} find negligible evolution in the metal content of massive galaxies at $0.1<z<0.7$ and ages consistent with passive evolution (see also \citealt{mendel2015}). 

Hydrodynamical simulations and semi-analytic models of galaxy formation and evolution are a powerful tool (see \citealt{somerville2015} for a review). In these models, galaxies are fed by cold flows of gas, stars, and dark matter which generate episodes of star formation (\citealt{keres2005}, \citealt{dekel2006}). When galaxies (or their dark matter haloes) reach a certain critical mass, a virial shock is created, slowing down the cooling rate and the galaxy star formation rate decreases. This is commonly referred to as ``mass quenching'' (\citealt{birnboim2007}) because the mechanism responsible for the decrease in star formation happens at a certain critical mass. Other physical processes invoked to quench galaxies are feedback events (e.g., AGN and stellar feedback; \citealt{croton2006}, \citealt{somerville2008}, \citealt{hopkins2014}), environmental effects (e.g., \citealt{peng2010}, \citealt{woo2013}), and morphological transformations (e.g., \citealt{gavazzi2015}, \citealt{zolotov2015}, \citealt{tacchella2016}).

By connecting observations of galaxies with predictions from simulations, we can attempt to understand the mechanisms that lead galaxies to quiescence. For example, \citet{porter2014} managed to reproduce the observed correlations between the stellar population parameters, age and metallicity, and the structural parameters, size and velocity dispersion, of early-type galaxies using a semi-analytic model based on the Bolshoi cosmological simulation. Other methods to compare simulations and observations are based on identifying the progenitors of galaxies and infer the processes involved in the transformation from one population to the other (\citealt{vandokkum2013}, \citealt{patel2013}, \citealt{papovich2015}). The strength of this technique is the ability to assess the evolution of galaxies in a model independent way. \citet{torrey2015} and \citet{terrazas2016} show, however, that the large variety of stellar and dark matter assembly histories in cosmological simulations cannot be captured by constant number density analyses.

Another method is to infer galaxy star formation histories from the fossil record of their past stellar populations (e.g., \citealt{panter2003}, \citealt{thomas2005}, \citealt{panter2007}, \citealt{gonzalez2014}, \citealt{mcdermid2015}, \citealt{citro2016}, \citealt{fumagalli2016}), sampling wide ranges of stellar masses and redshifts. For example, \cite{mcdermid2015} and \cite{citro2016} measure the SFHs of massive early-type galaxies and find that these systems are very old, forming the bulk of their mass in the first 1--2 Gyr. For our work, we choose this strategy, and we therefore need a large dataset that samples the variety of galaxies in the Universe and accurate models of their spectral energy distributions. We already used a similar approach in \citet{pacifici2016} where we assessed the SFHs of local ($z\sim0.07$) galaxies.

The Cosmic Assembly Near-IR Deep Extragalactic Legacy Survey (CANDELS; \citealt{grogin2011}, \citealt{koekemoer2011}) provides us with deep photometric data from the \textit{Hubble Space Telescope} (\textit{HST}) optical and near-infrared cameras (ACS and WFC3). These photometric data are carefully matched with ground-based observations from the $U$ to the $K$ band and with \textit{Spitzer}/IRAC photometry up to 8$\mu m$. Such a catalog is the optimal dataset to sample a large range in galaxy properties with sufficient statistics.

To interpret large, multi-wavelength datasets like CANDELS, we have developed a comprehensive spectral energy distribution (SED) fitting tool (\citealt{pacifici2012}) that relies on physically motivated models of galaxy formation and evolution. Instead of relying on simple parametric forms for galaxy star formation histories (SFHs) and assuming galaxies do not evolve in metallicity, as is generally done, our model is based on a combination of star-formation and chemical enrichment histories from hierarchical simulations of galaxy formation. Even though our model SEDs are based on such histories, our results are not biased by the particular model adopted. Such SFHs provide a large variety of possible galaxy evolutionary paths and thus the parameter space sampled by the models is much larger than traditional approaches. This is particularly evident when comparing model and observed colors of galaxies (\citealt{pacifici2015}). Furthermore, the assumption of an evolving metallicity is crucial to overcome the degeneracy between age and metallicity when interpreting the colors of galaxies.

In addition to adopting complex SFHs, our models incorporate nebular emission and state-of-the-art models of the spectral evolution of stars, gas, and dust, spanning a very wide range in the space of star formation rates (SFRs), metallicities, and dust and gas properties. They thus have the power to constrain the SFHs of different populations of galaxies at different redshifts. For example, assuming simple functional forms for the SFHs where the metallicity is fixed and the SFR can only decline with time does not allow one to sample properly the colors of all galaxies, especially the youngest ones, and thus correctly interpret their SEDs.

Key to distinguishing quiescent galaxies from dusty, star-forming ones is the treatment of dust attenuation in SED models. It has been shown that the attenuation by dust is more complex than the single attenuation law that is typically assumed (see for example \citealt{witt2000}, \citealt{chevallard2013}, and \citealt{salmon2015} for studies on different dust attenuation curves). Thus, building on the work by \citet{chevallard2013}, in our SED modeling, we implement a two-component dust model that spans a variety of attenuation laws.

Combining the CANDELS dataset and our new approach, we focus on red, quiescent galaxies and measure average SFHs as a function of stellar mass and redshift. We can thus answer questions such as: What are the average shapes of the SFHs of quiescent galaxies? Do these shapes change with redshift and stellar mass? Does the formation time of quiescent galaxies vary with stellar mass? At what time and at what stellar mass do galaxies stop forming stars? With these questions answered, we can then place important constraints on the main mechanisms involved in the formation and evolution of quiescent galaxies.

This paper is organised as follows. In Section~\ref{sec:data} we present the details of the observational dataset. The SED modelling approach, fitting procedure, and selection of quiescent galaxies are described in Section~\ref{sec:model}. In Section~\ref{sec:assembling} we show how we create median SFHs of galaxy populations as a function of stellar mass and redshift. In Section~\ref{sec:results}, we show that stellar mass and redshift regulate the quenching of star formation. In Section~\ref{sec:discussion}, we place the constraints on the mechanisms involved in the evolution of quiescent galaxies. We summarize our results in Section~\ref{sec:concl}.

Throughout the Paper, we adopt a Chabrier initial mass function (IMF, \citealt{chabrier2003}) and a standard $\Lambda$CDM cosmology with $\Omega_{\mathrm{M}}=0.3$, $\Omega_{\Lambda}=0.7$, $h=0.7$. Magnitudes are given in the AB system.

\section{Data}
\label{sec:data}

We use \textit{HST}/WFC3-F160W-selected catalogs from the CANDELS Survey (\citealt{grogin2011,koekemoer2011}) in the GOODS-South (\citealt{guo2013}) and GOODS-North (Barro et al., in preparation) fields. These fields are similar in terms of wavelength coverage and depth. They cover a combined area of 329 arcmin$^{2}$ on the sky. 

Multi-wavelength photometry is measured using TFIT, following \citet{guo2012}. The photometry includes a maximum of 17 and 18 measurements in GOODS-South and GOODS-North, respectively (we will discard galaxies with less than 10 measurements, see Section~\ref{sec:fitting}). These are from space-based and ground-based observations, sampling from $U$ to \textit{Spitzer}/IRAC. In GOODS-North, the catalog includes also observations in the \textit{HST}/WFC3-F275W band. These cover the central part of the field providing ultraviolet fluxes for 14\% of the galaxies. Filters and depths are listed in Table~\ref{tab:data}. Photometric redshifts are derived by combining the results of different photometric-redshift codes, as explained in \citet{dahlen2013}. These redshifts and their uncertainties are then used as priors by our SED-fitting code (see Section~\ref{sec:fitting}).

The total number of galaxies in the GOODS-South and GOODS-North catalogs are 34,930 and 35,445, respectively, totaling 70,375. We account for the `class star' and `no use' flags in the catalog and we are left with 64,175 galaxies. We select galaxies with $H<26$ (where $H$ is \textit{HST}/WFC3-F160W) to balance between acceptable signal-to-noise ratios ($S/N>5$) and the inclusion of faint galaxies. This reduces the number to 36,378 galaxies. We then select the galaxies in the redshift range $0.2<z<2.1$, which samples roughly 8 Gyr in the lifetime of the Universe. With this cut, we are left with 27,722 galaxies. We will discuss mass-completeness limits and the selection of quiescent galaxies in Section~\ref{sec:fitting} and \ref{sec:sample}, respectively.

{\renewcommand{\arraystretch}{1.2}
\begin{table}
\caption{Filters and depths for the observations in GOODS-South and GOODS-North.}
\label{tab:data}
\begin{center}
\begin{tabular}{l r | l r}
\hline
\hline
\multicolumn{2}{c}{GOODS-South$^{a}$}&\multicolumn{2}{c}{GOODS-North$^{b}$}\\
Filter & Depth (mag)  &Filter & Depth (mag)\\
\hline
		&			& WFC3-F275W&	29.5 \\
$U$ CTIO& 26.6		& $U$ KPNO&		26.9 \\
$U$ VIMOS& 28.0		& $U$ LBC& 		28.2	\\
ACS-F435W& 28.9		& ACS-F435W & 	28.0	\\
ACS-F606W& 29.3		& ACS-F606W& 	28.4	\\
ACS-F775W& 28.5		& ACS-F775W& 	28.1	\\
ACS-F814W& 28.8		& ACS-F814W& 	28.6	\\
ACS-F850LP& 28.5		& ACS-F850LP& 	28.0	\\
WFC3-F098M& 28.8		& WFC3-F105W& 	28.2	\\
WFC3-F105W& 27.4		& WFC3-F125W& 	28.5	\\
WFC3-F125W& 27.7		& WFC3-F140W& 	28.1	\\
WFC3-F160W& 27.4		& WFC3-F160W& 	27.5	\\
$Ks$ ISAAC & 25.1		& $K$ MOIRCS& 	26.0	\\
$Ks$ HAWK-I& 26.4		& $Ks$ CFHT& 	24.9	\\
$3.6\mu$m IRAC& 26.5	& $3.6\mu$m IRAC& 25.5	\\
$4.5\mu$m IRAC& 26.2	& $4.5\mu$m IRAC& 25.5	\\
$5.8\mu$m IRAC& 23.7	& $5.8\mu$m IRAC& 23.0	\\
$8.0\mu$m IRAC& 23.7	& $8.0\mu$m IRAC& 23.0	\\
\hline
\multicolumn{4}{l}{$^{a}$ \protect\citet{guo2013}}\\
\multicolumn{4}{l}{$^{b}$ Barro et al., in preparation}\\
\end{tabular}
\end{center}
\end{table} 
}

\section{Galaxy SED models and fits to the observations}
\label{sec:model}

{\renewcommand{\arraystretch}{1.2}
\begin{table*}
\caption{Prior distributions of the physical parameters in our SED models.} 
\begin{center}
\begin{tabular}{l c c}
\hline
\hline
Parameter & Description & Range\\
\hline
$z_{\mathrm{obs}}$ & Redshift of observation & $0.1<z_{\mathrm{obs}}<2.5$ \\
$t_{\mathrm{evo}}$ & Evolutionary stage (in lookback time) & $t_{\mathrm{obs}}<t_{\mathrm{evo}}<t_{\mathrm{obs}}+3$ Gyr \\
$\ssfr=\sfr/M_{\ast}$ & Specific star formation rate [when $\log(\ssfr/\mathrm{Gyr}^{-1})>-2$] & $-2<\log(\ssfr/\mathrm{Gyr}^{-1})< 2$ \\
$\log(O/H)$ & Gas-phase oxygen abundance (when $\sfr>0$) & $7<12+\log(O/H)<9.4$ \\
\tauv & Total $V$-band attenuation optical depth of the dust [when $\log(\ssfr/\mathrm{Gyr}^{-1})>-2$] & $0.01<\tauv<4$ \\
\tauv & Total $V$-band attenuation optical depth of the dust [when $\log(\ssfr/\mathrm{Gyr}^{-1})\leqslant-2$] & $0.01<\tauv<2$ \\
$\mu$ & Fraction of \tauv\ contributed by dust in diffuse ISM & $0.1<\mu<0.7$ \\
$n$ & Slope of the attenuation curve in the diffuse ISM & $-1.0<n<-0.4$ \\
\hline
\end{tabular}
\end{center}
\label{tab:params}
\end{table*}
}

\subsection{Library of model SEDs}
\label{sec:library}

We use the model spectral library based on the approach of \citet{pacifici2012} tailored to fit the dataset described in Section~\ref{sec:data}. Our model includes realistic star formation and chemical enrichment histories derived from a semi-analytic post-treatment (\citealt{delucia2007}) of the Millennium cosmological simulation (\citealt{springel2005}). This allows us to explore a large variety of SFH shapes, such as rising, declining, roughly constant, bursty and smooth. We generate a sample of 500,000 model galaxies, selecting randomly the redshift of observation ($z_{\mathrm{obs}}$) between 0.1 and 2.5, to appropriately sample the redshift range of the observed sample ($0.2<z<2.1$), and the evolutionary stage between the lookback time at the redshift of observation ($t_{\mathrm{obs}}$) and $t_{\mathrm{obs}} + 3$ Gyr (see Sections 2.1 and 3.1.2 of \citealt{pacifici2012}). This procedure allows us to include galaxies that start forming stars at different epochs. We also resample the current (i.e., averaged over the last 10 Myr) specific SFR (which is defined as SFR/$M^{\ast}$ and we demarcate as sSFR or $\psi_S$) between 0.01 and 100 Gyr$^{-1}$, and the current gas-phase metallicity between $7<12+\log(O/H)<9.4$. This current metallicity affects the nebular emission and the stellar populations younger than 10Myr. Practically, it is a nuisance parameter because it cannot be inferred from photometry alone.

To create SEDs for each model galaxy, we compute the emission by the stars using the latest version of the \citet{bruzual2003} stellar population synthesis models. We process the nebular emission (both lines and continuum) consistently with the stellar emission by running the photoionization code CLOUDY (\citealt{ferland1996}) on the simple stellar populations of ages less than 10 Myr (the nebular emission at older ages is negligible) and we assume that galaxies are ionization bounded (see \citealt{charlot2001} for details).

We model the attenuation by dust by applying the two-component prescription following \citet{charlot2000} as in \citet{pacifici2012} (see also \citealt{chevallard2013}). We draw randomly the slope of the attenuation curve due to the interstellar medium (ISM) between -1.0 and -0.4. The slope of the attenuation curve due to the birth clouds (BC) surrounding young stars is fixed to -1.3. We vary the relative proportion of dust in the BC and in the ISM ($\mu=\tau_{V}^{ISM}/\tau_{V}$, where $\tau_{V}=\tau_{V}^{ISM}+\tau_{V}^{BC}$) in the range $0.1<\mu<0.7$. A summary of the prior distributions of key physical parameters is given in Table~\ref{tab:params}. The priors are globally flat. The distributions of sSFR and $\tau_{V}$ smoothly declines at both edges and at the upper edge, respectively.

Hence, we have high-resolution SEDs for all the galaxies in the model spectral library. We than apply photometric filter functions used for the observations to produce photometry from the high-resolution model SEDs.

\subsection{Fitting procedure}
\label{sec:fitting}

\begin{figure}
\begin{center}
\includegraphics[width=0.47\textwidth]{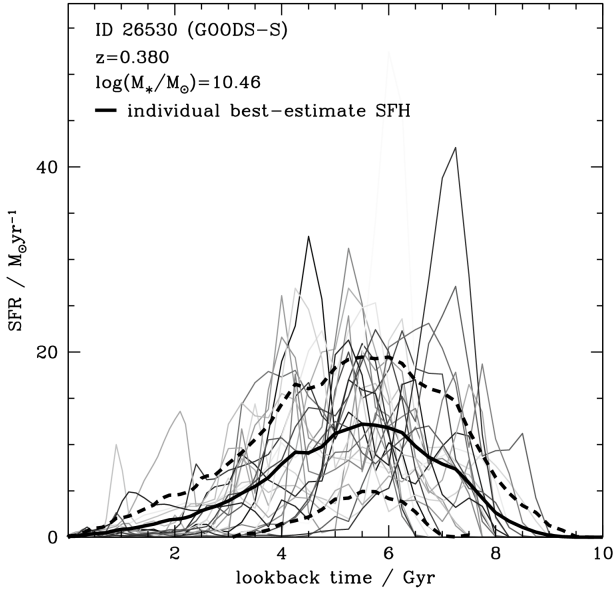}
\caption{The best-estimate SFH of a single galaxy is the average of all the models weighted by their likelihoods. We show the best-estimate SFH of an example galaxy as a thick black line. The standard deviation about the average is represented by the thick dashed line. The first 25 best-fit models for this galaxy are shown as thin lines of different shades of gray where the darker the shade, the higher the likelihood. Note that a lookback time of 0 corresponds to the lookback time at the redshift of observation.}
\label{fig:10sfh}
\end{center}
\end{figure}

\begin{figure*}
\begin{center}
\includegraphics[width=0.9\textwidth]{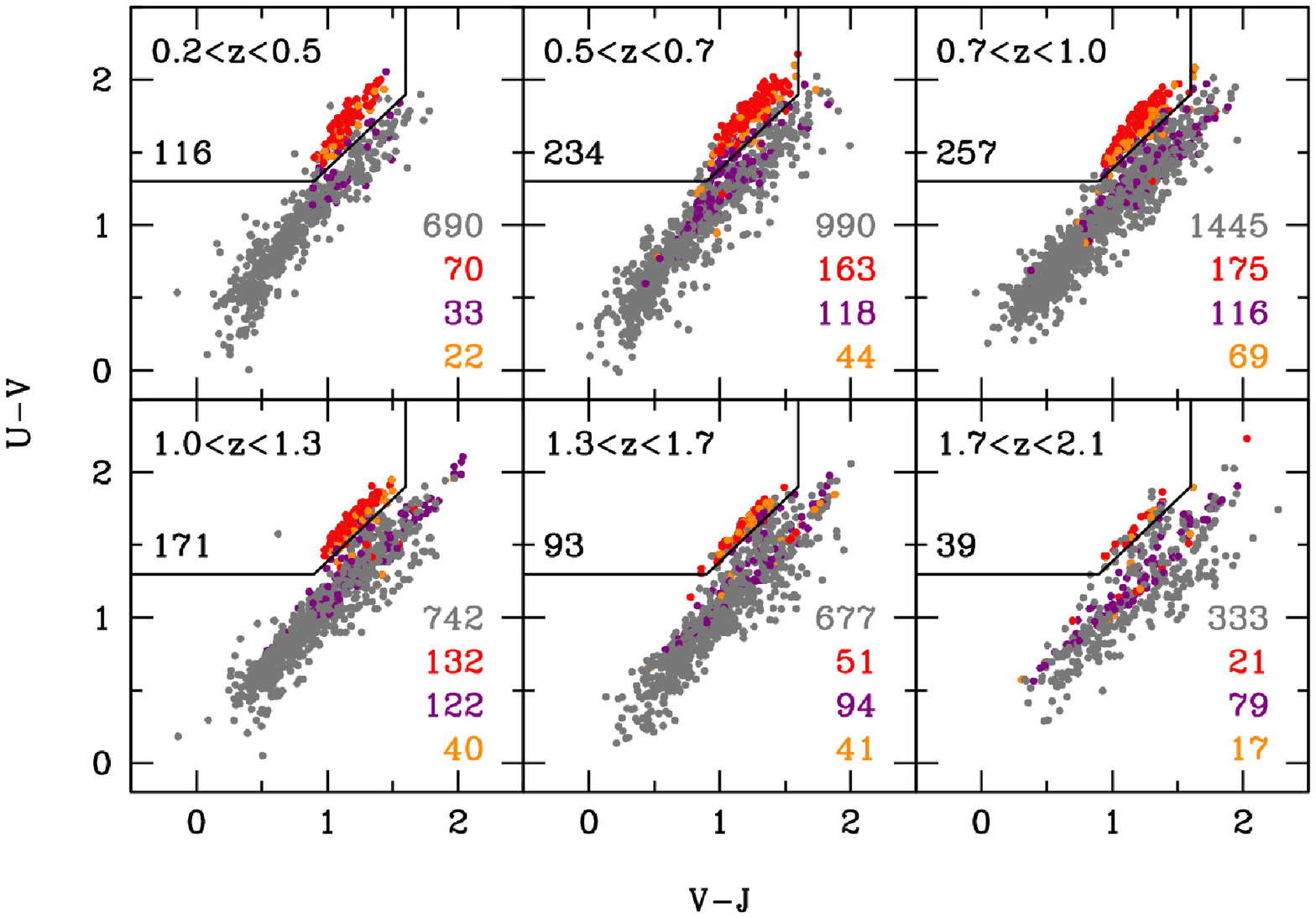}
\caption{We select quiescent galaxies applying a threshold in sSFR and we test that this selection is consistent with the commonly used selection in $UVJ$ colors. In the $U-V$ versus $V-J$ diagram, quiescent galaxies occupy the area in the upper left corner. The vast majority of quiescent galaxies selected with our sSFR criterion (red points) lie within this box. Star-forming galaxies are marked in gray. We also demarcate galaxies which are quiescent (star-forming) according to our sSFR criterion, but due to the uncertainties in their sSFRs, could be classified as star-forming (quiescent) as orange (purple) points.  Using this color-color selection or excluding the grey and purple data in our analysis does not qualitatively change our results. The number of galaxies in each category is reported in each panel. The number of galaxies which are quiescent according to the color-color criterion are provided in the upper left box.}
\label{fig:uvj}
\end{center}
\end{figure*}

We use the library of 500,000 model SEDs described in Section~\ref{sec:library} to fit the multi-band photometry of the galaxies in the observational sample described in Section~\ref{sec:data}. For each galaxy, we use the photometric redshift from the catalog as prior (we assume a flat prior within $\pm1\sigma$).  We do not include the contamination by dust \textit{emission} in our models, therefore we fit only to the bands that sample rest-frame wavelengths shorter than $3 \mu m$. This means that for galaxies at $z<0.34$, we do not fit the \textit{Spitzer}/IRAC bands. If a galaxy is detected at $0.34<z<0.67$, we fit up to \textit{Spitzer}/IRAC channel 1. At $0.67<z<1.17$, we include \textit{Spitzer}/IRAC channel 1 and 2, and at $1.17<z<2.1$, we include \textit{Spitzer}/IRAC channel 1, 2, and 3. We adopt a minimum photometric error of 5\% to enlarge the number of models contributing to each fit and account for possible systematics in the photometric measurements.

We use the Bayesian approach described in \citet{pacifici2012} to extract, for each observed galaxy, probability density functions of stellar mass ($M^{\ast}$), star formation rate (which we demarcate as SFR or $\psi$), and $U-V$ and $V-J$ rest-frame colors. We also re-derive a value of redshift ($z$) within the prior so that it is consistent with all the other derived physical parameters. Typical uncertainties for these parameters are 0.1 dex,  0.3--0.5 dex, 0.05 mag, 0.05 mag, and 0.02--0.05, respectively. To make sure that the catalog redshifts (obtained with different SED modeling prescriptions) are not introducing biases in our results, we perform a test: we fit the photometry of the galaxies in the sample leaving the redshift as a free parameter. We find that the physical parameters derived with and without a prior on the redshift agree within their uncertainties for 98\% of the galaxies. We expect such good agreement because the redshifts in the catalog are derived by combining different approaches which assume different SED modeling. Potential biases are thus smaller than the uncertainties which we use as prior.

For each galaxy in the observational sample, we compute the best-estimate SFH (SFR as a function of lookback time) by averaging all the SFHs in the model library weighted by their likelihoods (in practice, we compute a likelihood-weighted average of the SFR at each lookback time). Sampling a large wavelength range in the observations allows us to distinguish among different effective ages, and thus different SFHs. Galaxies with similar effective ages and different SFHs are slightly more difficult to distinguish simply from broad-band colors. We will consider this caveat in the discussion section when interpreting our results. Figure~\ref{fig:10sfh} shows the best-estimate SFH (average and standard deviation) of an example observed quiescent galaxy. The SFHs that contribute to the average have a variety of shapes. By using a weighted average, we account for possible degeneracies in age, metallicity, and dust attenuation in a statistical way. We see in this example that all models contributing significantly to the average are quiescent at the time of observation and form the bulk of their stars in a similar window of time. This suggests that such degeneracies are not too severe. We verify that this is the case for most of the galaxies included in the final sample.

We consider the quality of the fits and the mass completeness limits. We flag galaxies that are observed in less than 10 photometric bands and that have $\chi^2/N>5$,\footnote{The distribution of $\chi^2/N$ peaks around 1 with a short tail that extends above 5.} where $\chi^2$ is the chi-square of the best-fit model and $N$ is the number of bands included in the fit. With the first criterion, we remove 1018 galaxies. With the second criterion, we remove 1929 galaxies of which 205 are in GOODS-South and 1724 are in GOODS-North. The `no use' flags in GOODS-North have not been finalized yet (Barro et al. in prep), thus we expect to find additional galaxies with bad fits (in case of contamination by star spikes, contamination by neighboring galaxies, or too little uncertainties in some photometric bands). Our criterion helps us flagging out such objects. We also compare the observed colors with the colors of the model spectral library and we find a good match (as in \citealt{pacifici2015}). The majority of the galaxies we flag show blue UV-optical colors and thus do not affect our selection of quiescent galaxies anyway (see Section~\ref{sec:sample}). We note that even with 10 bands, the full wavelength range is well sampled because at least a $U$ band, the \textit{HST}/ACS bands, two \textit{HST}/WFC3 bands, a $Ks$ band, and \textit{Spitzer}/IRAC channel 1 are measured for most galaxies.

In the stellar-mass range $8.75<\log(M_{\ast}/M_{\odot})<11.75$, we count 12,594 galaxies.\footnote{We note that, if we do not perform any magnitude cut, we are left with the same sample, because 99\% of the galaxies with stellar mass larger than $\log(M_{\ast}/M_{\odot})=8.75$ at the redshifts considered in this work are detected at $H<26$. The residual 1\% is composed exclusively of star forming galaxies.} We divide these galaxies into six redshift bins ($0.2<z<0.5$, $0.5<z<0.7$, $0.7<z<1.0$, $1.0<z<1.3$, $1.3<z<1.7$, and $1.7<z<2.1$), chosen so that the number of galaxies in each bin is similar. The mass completeness limits (to $H<26$) in the six redshift bins are $\log(M/M_{\odot})=8.4, 8.7, 9.1, 9.5, 9.7, 10$, respectively. Removing galaxies below the completeness limits, we are left with 6284 galaxies.

\subsection{Selecting quiescent galaxies in the observations}
\label{sec:sample}

Quiescent galaxies are commonly identified in two different ways: setting a threshold in sSFR and thresholds in rest-frame colors (the $UVJ$ diagram). We show here that these two methods yield consistent results and we adopt the former technique. We obtain the same qualitative results if we adopt the color-color selection technique.

We first measure SFRs and stellar masses of the galaxies in the observational sample by fitting our SED models to their photometry (Section~\ref{sec:fitting}). On the SFR-$M^{\ast}$ plane, star-forming galaxies occupy a well defined sequence (e.g., \citealt{noeske2007}, \citealt{whitaker2012}). Quiescent galaxies fall below this sequence. We can thus define a threshold in sSFR (SFR/$M^{\ast}$) to distinguish quiescent from star-forming galaxies. The normalization of the star-formation main sequence (i.e., the SFR-stellar mass relation; see for example \citealt{noeske2007} and \citealt{whitaker2012}) evolves with redshift (as shown, for example, by \citealt{whitaker2012} and \citealt{fumagalli2014}), thus also the threshold between star-forming and quiescent galaxies evolves in the same way. We parametrize the evolution of the threshold as $\psi_S(z) = 0.2/t_{U}(z)$, where $\psi_S(z)$ [Gyr$^{-1}$]  and $t_U(z)$ [Gyr] are the sSFR and the age of the Universe at redshift $z$, respectively. With this selection, we identify 845 quiescent galaxies. This is our final sample that is used in the rest of the paper.

Now, we compare the threshold in sSFR to the threshold in color-color space. In Figure~\ref{fig:uvj}, we show the $U-V$ and $V-J$ rest-frame colors of the 6284 galaxies in the full observational sample (this includes both star-forming and quiescent galaxies), in the six redshift bins. At all redshifts, quiescent galaxies selected in sSFR are well confined to the color-color region used to select quiescent galaxies (\citealt{williams2009}). Moving from low to high redshifts, quiescent galaxies become progressively younger and thus lie closer to the diagonal edge of the box (\citealt{whitaker2012}; Fang et al., in preparation). At low redshifts, a $UVJ$ cut would select slightly more quiescent galaxies than the selection in sSFR, while at redshift $z>0.7$ the two selection criteria are roughly equivalent in numbers. In the highest redshift bins the uncertainties in the retrieved rest-frame colors and sSFRs increase, but the selection remains acceptable. The selection in $UVJ$ colors is thus equivalent to our selection in sSFR. The qualitative results described in Section~\ref{sec:assembling} do not change if quiescent galaxies are selected from the $UVJ$ diagram instead of being selected in sSFR. This derives from the fact that the valley in between the quiescent and the star-forming regions in the $UVJ$ diagram is not heavily populated. Also it suggests that the trends in SFH are smooth moving from the quiescent to the star-forming region. We are exploring these trends as part of a separate work where we assess the SFHs of both quiescent and star-forming galaxies.

To summarize the sample selection, we started with 70,375 galaxies in the two fields. We selected galaxies with good flags in the catalog and $H<26$ (36,378 galaxies). We selected the redshift range $0.2<z<2.1$ (27,722), and the stellar mass range $8.75<\log(M_{\ast}/M_{\odot})<11.75$ (12,594). We exclude 1929 galaxies with problematic photometry. We then remove galaxies below the completeness limits and we are left with 6284 galaxies. Of these, 845 are quiescent.

\section{Median SFHs as a function of stellar mass and redshift}
\label{sec:assembling}

\begin{figure*}
\begin{center}
\includegraphics[width=0.47\textwidth]{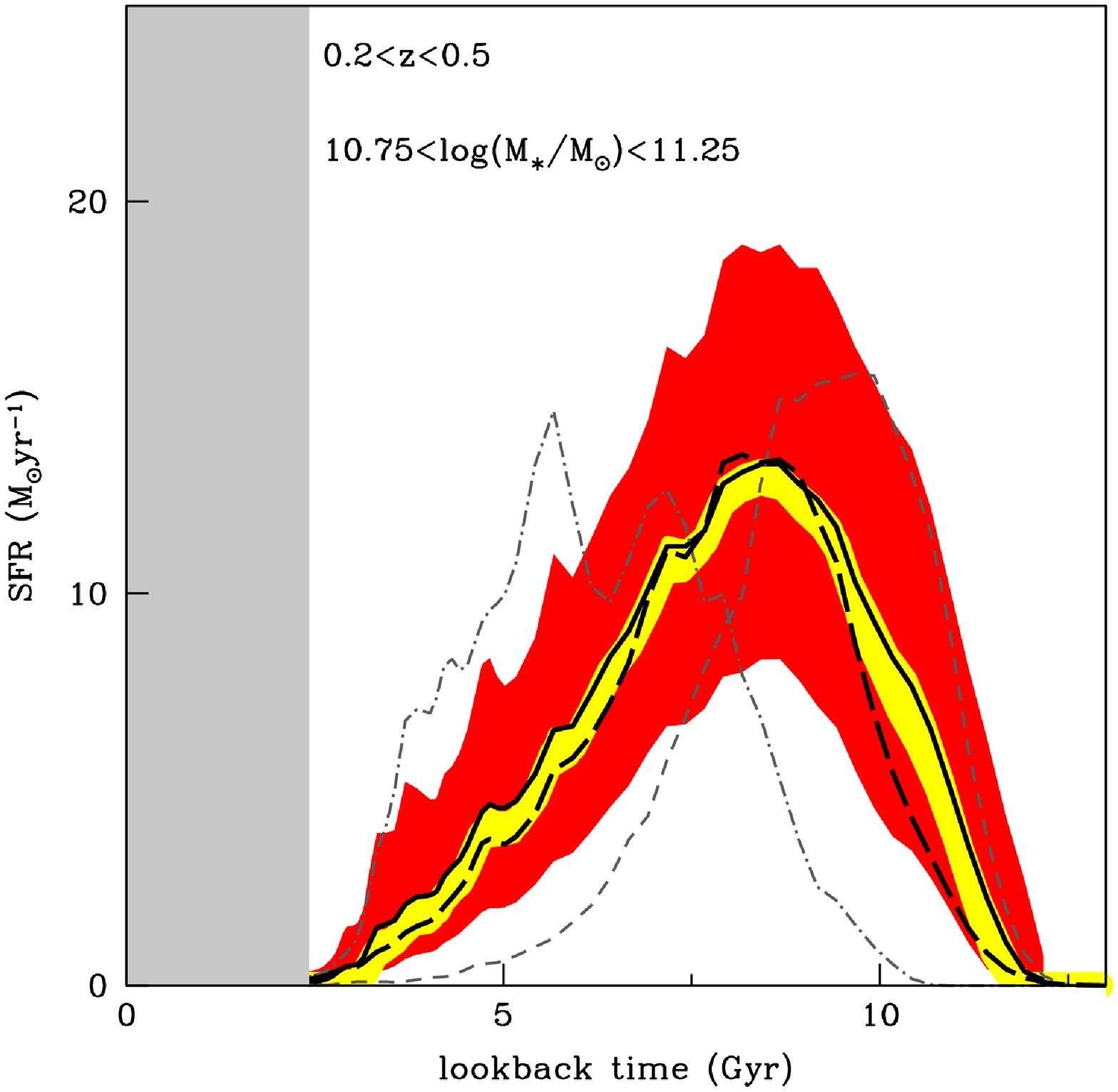}
\includegraphics[width=0.47\textwidth]{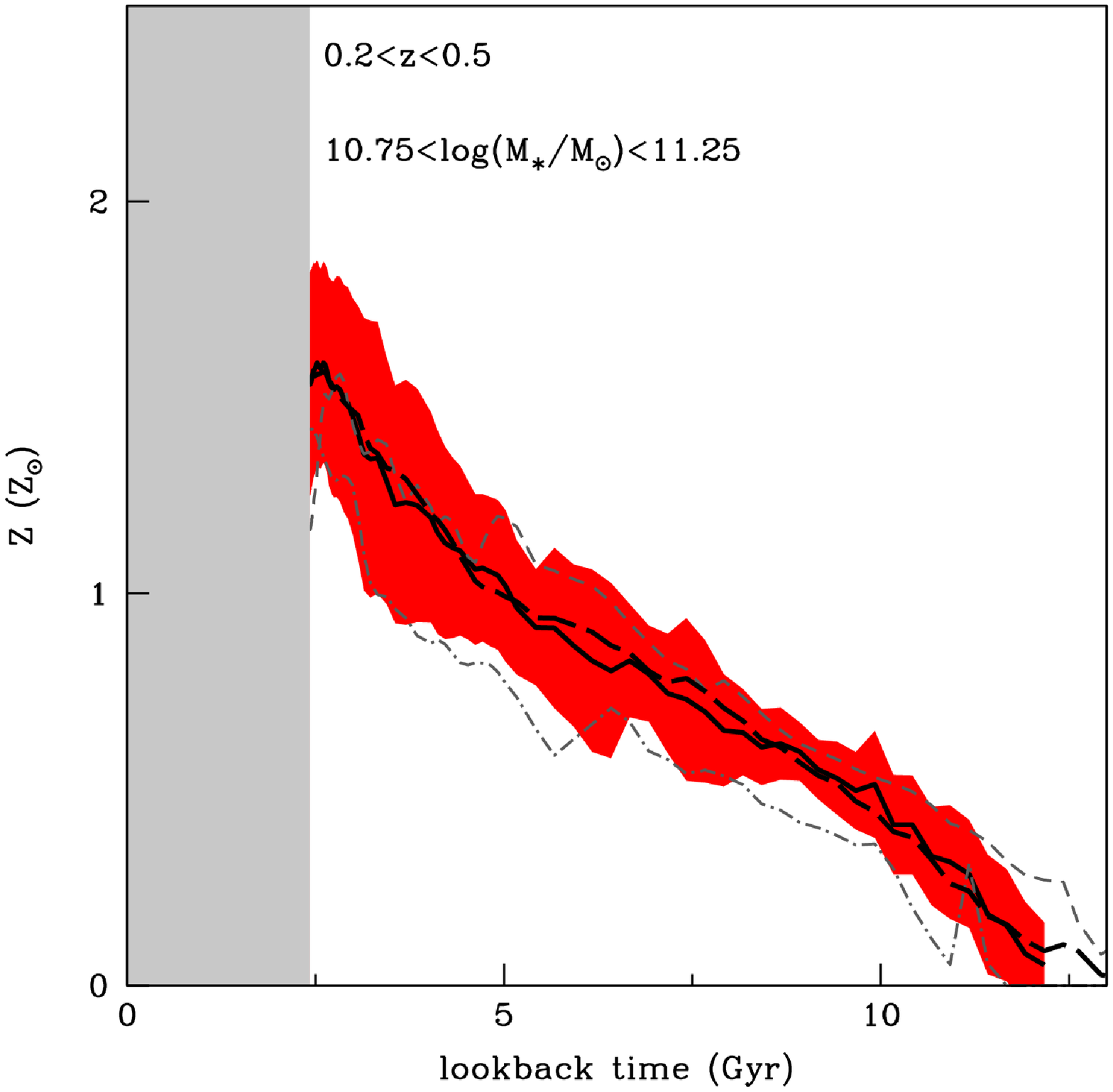}
\caption{The calculation of the median SFH (left-hand side) and metal enrichment history (right-hand side) for 24 galaxies in an example bin of stellar mass and redshift is demonstrated. The medians are shown as thick black lines, averages as long-dashed black lines, and two examples of individual galaxies are shown as thin dashed and dot-dashed grey lines (the other 22 are not shown for clarity). The interquartile range (25th to 75th percentile) is shown as a red shade. The yellow shade on the left-hand side represents the sample variance, which is the confidence range about the medians obtained by bootstrapping 100 times the galaxies in this bin. For all galaxies a lookback time of 0 corresponds to $z=0$. Lookback times younger than the lookback time that corresponds to the observed redshift range are shaded grey.}
\label{fig:sfhsing}
\end{center}
\end{figure*}

\begin{figure}
\begin{center}
\includegraphics[width=0.47\textwidth]{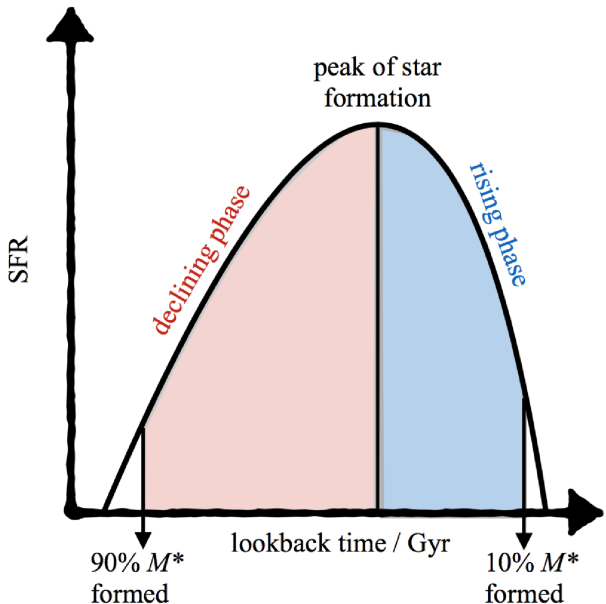}
\caption{Cartoon representation of how we quantify the characteristics of the median SFHs. Median SFHs rise, reach a peak, and decline. We measure when they peak and how long they take to rise and decline. We define the rising phase to be the time between when 10\% of the total stellar mass is formed and the peak. We define the declining or quenching phase to be the time between the peak and when 90\% of the total stellar mass is formed.}
\label{fig:cartoonphases}
\end{center}
\end{figure}

\begin{figure*}
\begin{center}
\includegraphics[width=1\textwidth]{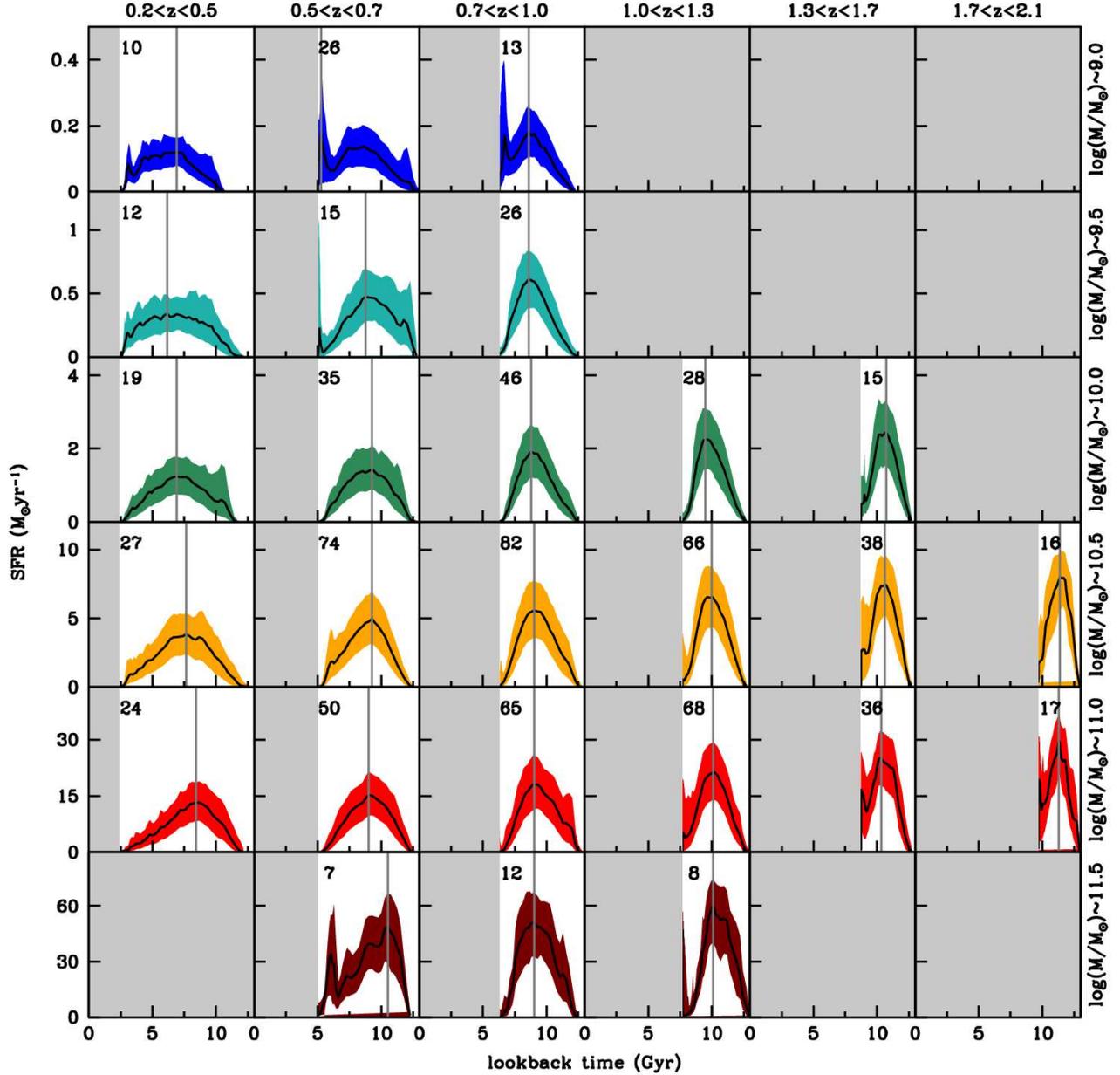}
\caption{Median SFHs of quiescent galaxies in bins of stellar mass (rows) and redshift (columns). Medians are shown as black lines, and confidence ranges (25-to-75 percentile of the distribution) are shown by the colored shadings. In each panel, the time range that is not observable given the age of the Universe is shaded grey. The number of galaxies contributing to median is reported in each panel. We do not show median SFHs in bins where the galaxies have stellar masses below the completeness limits and where the number of galaxies in the bin is less than 5. The vertical grey lines mark the peak of the SFR in each median SFH.}
\label{fig:sfh}
\end{center}
\end{figure*}

\begin{figure*}
\begin{center}
\includegraphics[width=0.47\textwidth]{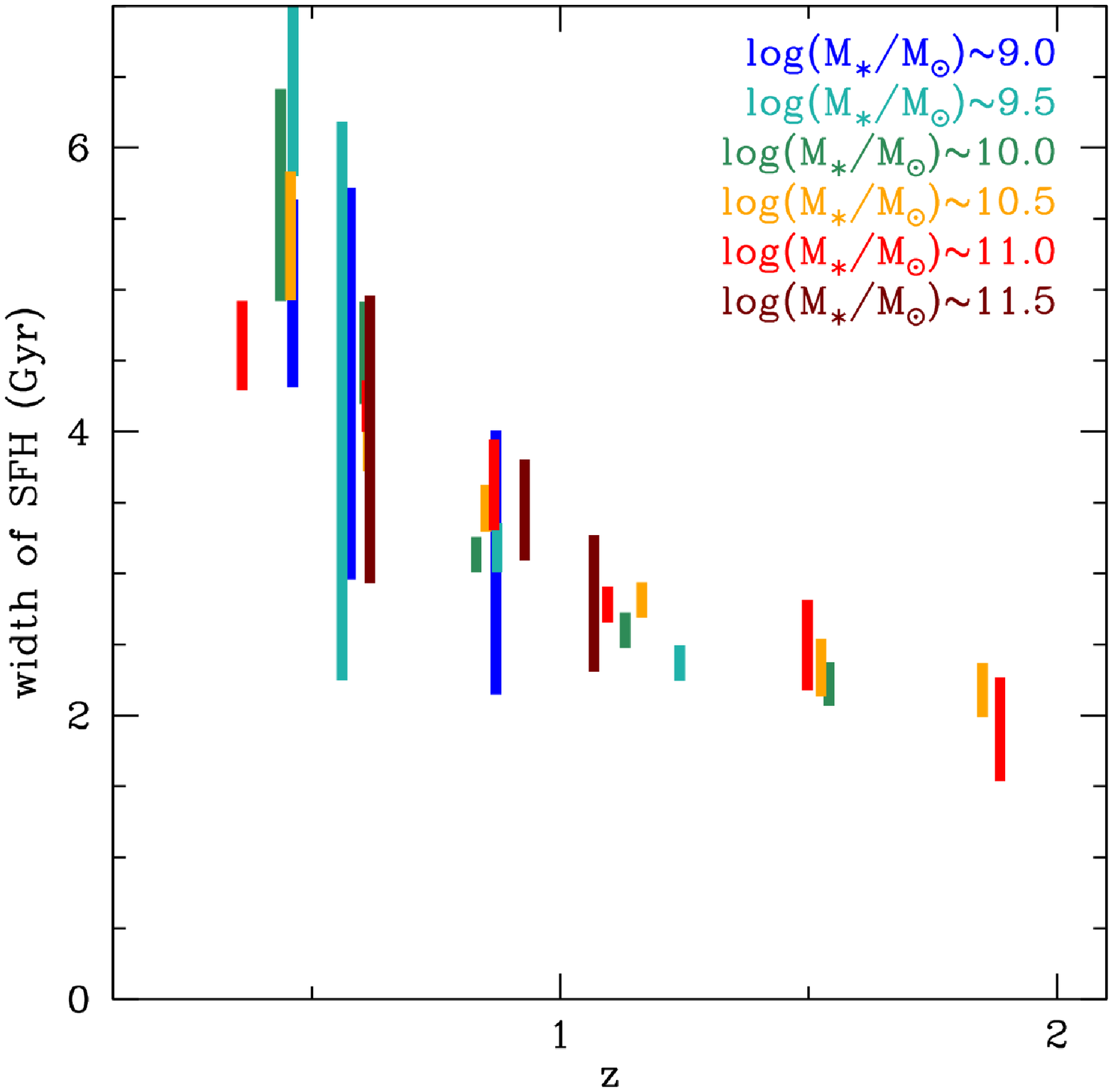}
\includegraphics[width=0.47\textwidth]{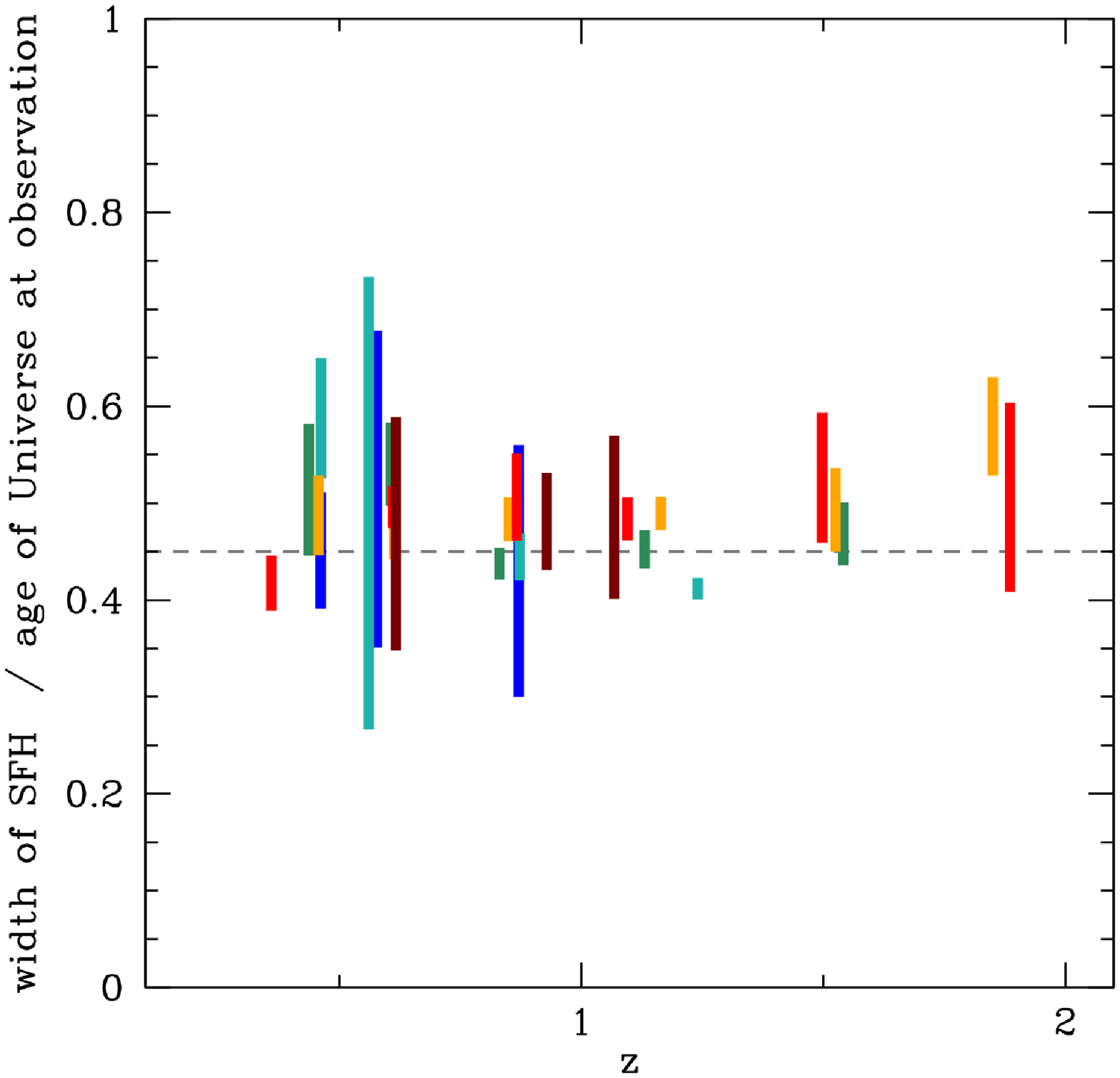}
\caption{The widths of the median SFHs in Figure~\ref{fig:sfh}, or the time galaxies spend in their star-forming phase, is a strong function of redshift but does not depend on stellar mass. The plot on the left-hand side shows the 25-to-75 percentiles of the distribution of the width of the median SFHs as a function of redshift. On the right-hand side we normalize the width to the age of the Universe at the redshifts of observation. The dashed grey line guides the eye to the value that best represents the observations (0.45). Each mass bin is color coded according to the scheme in Figure~\ref{fig:sfh}.}
\label{fig:width}
\end{center}
\end{figure*}

As described in Section~\ref{sec:model}, we fit our SED models to the sample of 845 quiescent galaxies and we derive a best-estimate SFH (average and standard deviation of all the models weighted by their likelihoods) for each. In this section, we compute median SFHs in bins of stellar mass and redshift. We then explore trends of these median SFHs. We find that if we use average SFHs instead of median SFHs, we obtain comparable results.

We divide the sample into six redshift bins, and, for each, we further divide the galaxies into six stellar-mass bins. We create median SFHs in each of the 36 bins, by stacking all the probability functions of the galaxies at each step in lookback time. We show an example bin in Figure~\ref{fig:sfhsing} (left-hand side). We calculate the distribution of the SFR of all galaxies in the bin at steps in lookback time (we set lookback time $t=0$ to be $z=0$). From the SFR distribution at each lookback time, we compute the median and interquartile range (25-to-75 percentile, which we refer to as the confidence range). The confidence range here represents the combination of the uncertainty on the single SFHs and the sample variance. To identify which of the two components dominates, we measure the sample variance by bootstrapping the galaxies in each bin and recalculating the median SFH. We do this 100 times and we measure the confidence range of the distribution of the medians. We find that the sample variance (shown in yellow in Figure~\ref{fig:sfhsing}) is in this case smaller than the uncertainty.

In the same way, we compute medians of the metal enrichment histories in the same 36 bins. In Figure~\ref{fig:sfhsing} (right-hand side), we show the median metal enrichment history for the same example bin as for the median SFH. The semi-analytic model from which we adopt the SFHs (see Section~\ref{sec:library}) predicts generally rising metal enrichment histories, consistent with observations of the evolution of the mass-metallicity relation with time (e.g., \citealt{lamareille2009} and \citealt{zahid2013}). The median metal enrichment histories we measure from the observations are likewise rising at all stellar masses and all redshifts. They show slightly lower metallicities at low stellar masses than at high stellar masses. This prior from the semi-analytic models allows us to address the age-metallicity degeneracy without using observations of spectral absorption lines (which is the typical way of addressing this issue; e.g., \citealt{worthey1992}, \citealt{faber1995}, \citealt{trager2000}).

We calculate median SFHs for all bins and we look for trends in their shapes which are quasi bell-shaped: the SFHs first rise, then reach a peak, and then decline (Figure~\ref{fig:cartoonphases}). We define the \text{rising phase} to be the time between when 10\% of the total stellar mass is formed and the peak. We define the \text{declining or quenching phase} to be the time between the peak and when 90\% of the total stellar mass is formed. In what follows, we refer to a duration roughly $\lesssim 2$ Gyr as \textit{fast}, and longer timescales ($\gtrsim 3$ Gyr) as \textit{slow}. These are absolute timescales and ignore the evolution of the dynamical time of galaxies. We will return to this point in the discussion. We remind the reader that we define a galaxy to be \textit{quenched or quiescent} when its sSFR goes below $0.2/t_{U}(z)$, where $t_U(z)$ [Gyr] is the age of the Universe at redshift $z$.

In Figure~\ref{fig:sfh}, we show the median SFHs of the quiescent galaxies in the sample in all bins. Here we qualitatively describe the results, and defer a quantitative analysis to the following section. The SFHs are quasi bell shaped because we are selecting only galaxies that are quenched at the time of observation, thus their SFR has to decrease before observation. At high redshift, when the Universe is young, both the rising and declining phases must happen fast (less than two billion years). At low redshift, these phases could be either short or long, because the Universe is old and thus more time is available. In two low-mass bins, we observe a sudden burst of star formation in the last billion year prior to observation. This is caused by a few ``post-starburst'' galaxies. Such galaxies almost quenched their star formation and then rejuvenated for a very short period of time before observation. Spectroscopic observations of these galaxies would confirm their nature (see for example \citealt{wild2016}).

We observe that low-mass galaxies take longer to reach their peak of star formation compared to high-mass galaxies. This means that for low-mass galaxies the rising phase is slower than for high-mass galaxies. Given that low-mass galaxies reach their peak late, their declining phase must happen fast. Conversely, high-mass galaxies reach their peak early, thus the declining phase can be either fast or slow. We observe it to be fast at high redshift (as already mentioned), and slow at low redshift. The difference between low- and high-mass galaxies in their declining phases suggests that different quenching mechanisms (that operate on different timescales) are involved.

The galaxies that are already quenched at high redshift are expected to be part of the low-redshift bin (provided that they do not ignite star formation again). Thus, we can calculate the upper limit on the number of galaxies that are quenched already at $z=2$ and are detectable in the bin at $0.2<z<0.5$. By calculating the densities of quiescent galaxies in the highest and the lowest redshift bins, we find that only about 13\% of the galaxies that are quiescent at low redshift are those that were already quiescent at $z=2$ (we indeed observe a few individual early quenched galaxies in the low-redshift high-mass bins, although the number statistics is too low to be able to confirm the 13\%). The remaining of the population of low-redshift quiescent galaxies must thus be formed by galaxies that have quenched their star formation later than $z=2$. This means that the population of quiescent galaxies is fed by star-forming galaxies that shut off their star formation at different epochs. In a separate work, we will combine this sample with a larger sample of low-redshift galaxies to investigate how many galaxies stay quiescent and how many rejuvenate.

We verify that the trends we observe are not simply a consequence of the prior of SFHs we are assuming (described in Section~\ref{sec:library}). To do so, we stack all the SFHs of the galaxies in the model library with SFR below the quiescent limit, using the same specific SFR threshold we apply to the data. We find that, at low redshift, the SFH prior is similar to a Gaussian function, but the confidence ranges are much larger than those observed in Figure~\ref{fig:sfh}. Also, stellar mass is only a scaling factor, thus, at a particular redshift, the same prior applies to all stellar mass bins. At high-redshift, the SFH prior of quiescent galaxies includes many galaxies with fast truncation. The stack of all quiescent model SFHs is a rising function with a fast truncation, which is very different from the median SFHs we observe in Figure~\ref{fig:sfh}. This test suggests that the trends we observe are data driven.

Summarizing, qualitatively low-mass galaxies increase slowly their star formation and then quench fast. High-mass galaxies instead reach their peak of star formation fast. They then quench fast at high redshift, when the Universe is young, and slow at low redshift. For future applications, we parametrize the median SFHs using analytic functions (\citealt{behroozi2013}) which seem to reasonably reproduce the shapes we observe. In Appendix~A, we report the best parameters for two plausible analytic functions at each redshift and stellar mass bin. These values can be used to model the SFHs of average populations of quiescent galaxies when more realistic prescriptions from cosmological simulations are not available.

\section{Properties of median SFHs}
\label{sec:results}

The median SFHs of quiescent galaxies are diverse. Galaxies can form stars and quench their star formation fast. Alternatively, they can form stars at a low rate for long periods of time and quench fast. They can also form the bulk of the mass at early times and then take a long time to reach quiescence. In the previous section, we demonstrated that the timescales of the rising and declining phases of SFHs depend on stellar mass and redshift. Here, we quantify these dependences. We measure from the median SFHs in Figure~\ref{fig:sfh} the following quantities: their width (i.e. when galaxies form the bulk of their stellar mass), when they reach the peak, how long they take to reach the peak, and how long they take to quench. We describe the trends and discuss their implications in Section~\ref{sec:discussion}.

\subsection{The widths of the SFHs}
\label{sec:width}

To quantify the time galaxies take to form the bulk of their stellar mass, we measure the full width at half maximum of the median SFHs in Figure~\ref{fig:sfh}. To do this, we bootstrap the galaxies in each bin and we calculate the full width at half maximum (FWHM) of the median SFHs 100 times. We then measure the 25--50--75 percentiles of the distribution of the 100 widths. This allows us to measure the sample variance in each bin. The actual uncertainty on the lookback time is roughly 10--20\% of the lookback time itself as shown by \cite{pacifici2016} (see their Figure 3 and Section 3.2).

Figure~\ref{fig:width} shows the 25-to-75 percentiles of the distribution of the widths as a function of redshift and stellar mass. The width are found to decrease with redshift. This is related to the time available to form stars at each redshift. In other words, at high-redshift, the SFHs of quiescent galaxies can only be narrow. In contrast, at low redshift, they can span the range from narrow to large, and we observe them to be large. We do not find a trend between width and stellar mass. Interestingly, when the widths of the SFHs are expressed as a fraction of the age of the Universe at the redshift of observation, we find a roughly constant value of 0.4--0.5, independent of redshift and stellar mass. This could be a fundamental result and we will address its implications in the discussion.

\subsection{Locations of the peaks of the SFHs}
\label{sec:peak}

\begin{figure*}
\begin{center}
\includegraphics[width=0.47\textwidth]{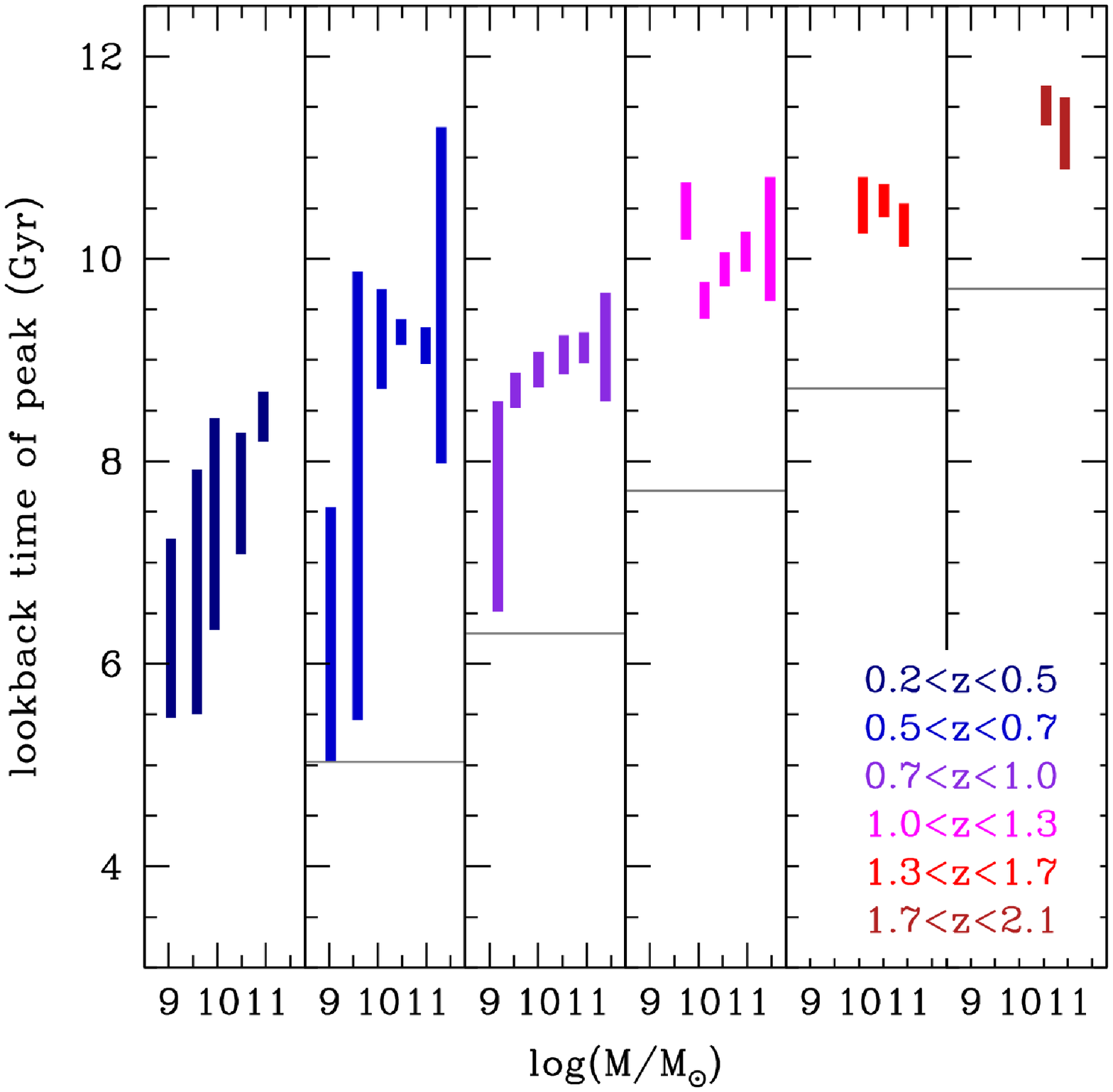}
\includegraphics[width=0.47\textwidth]{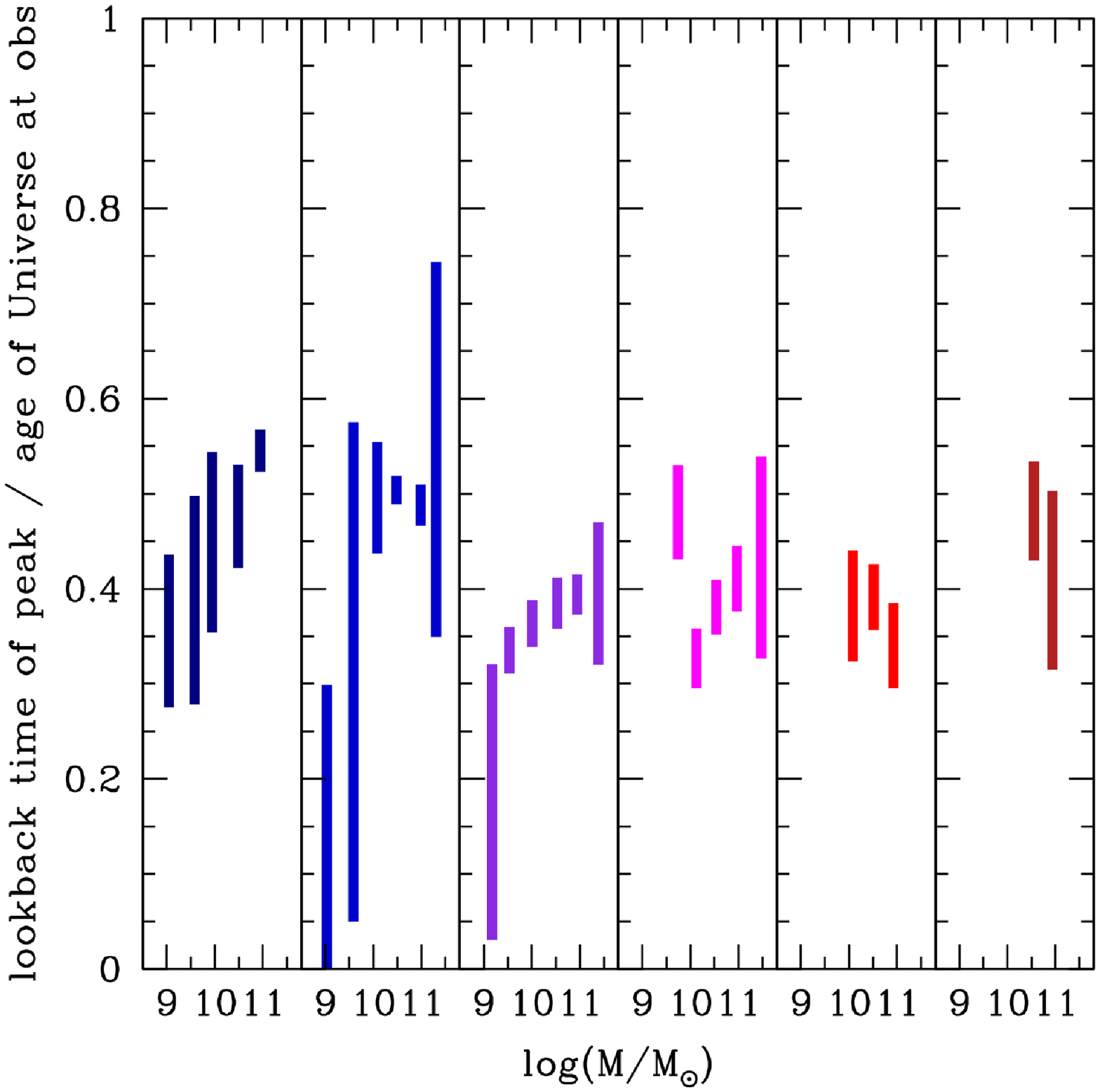}
\caption{The time at which galaxies reach the peak of their SFHs (vertical lines in Figure~\ref{fig:sfh}) is a function of their stellar mass and redshift. Each small panel represents a different redshift bin. Bars are color coded according to redshift. The horizontal gray lines mark the lookback time corresponding to the median redshift of observation of each bin (in the leftmost panel, the gray line is at 2.4 Gyr, thus below the y-axis limit). On the right-hand side, we normalize the lookback time of the peak to the age of the Universe at the redshifts of observation.}
\label{fig:peak}
\end{center}
\end{figure*}

\begin{figure*}
\begin{center}
\includegraphics[width=0.47\textwidth]{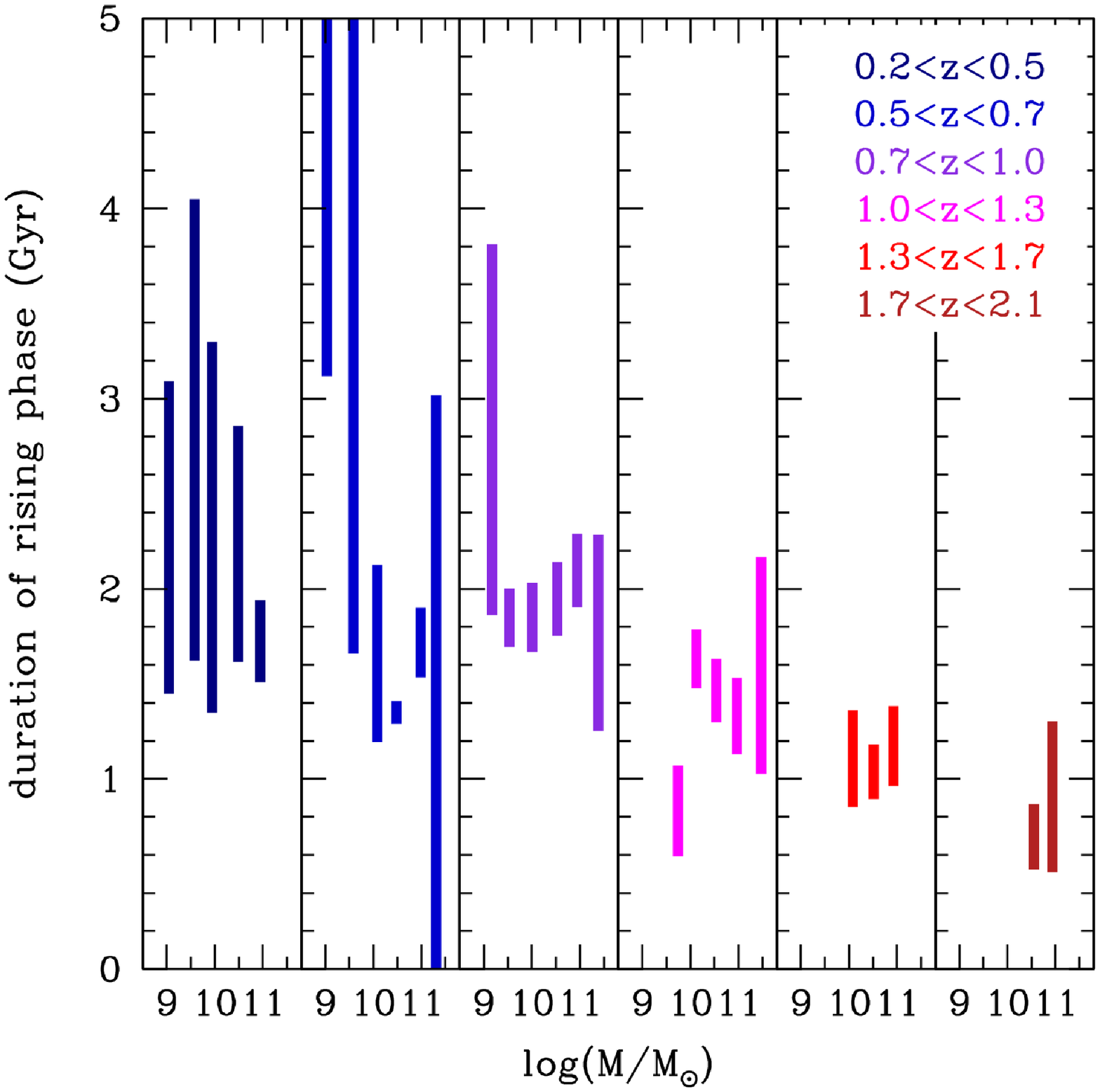}
\includegraphics[width=0.47\textwidth]{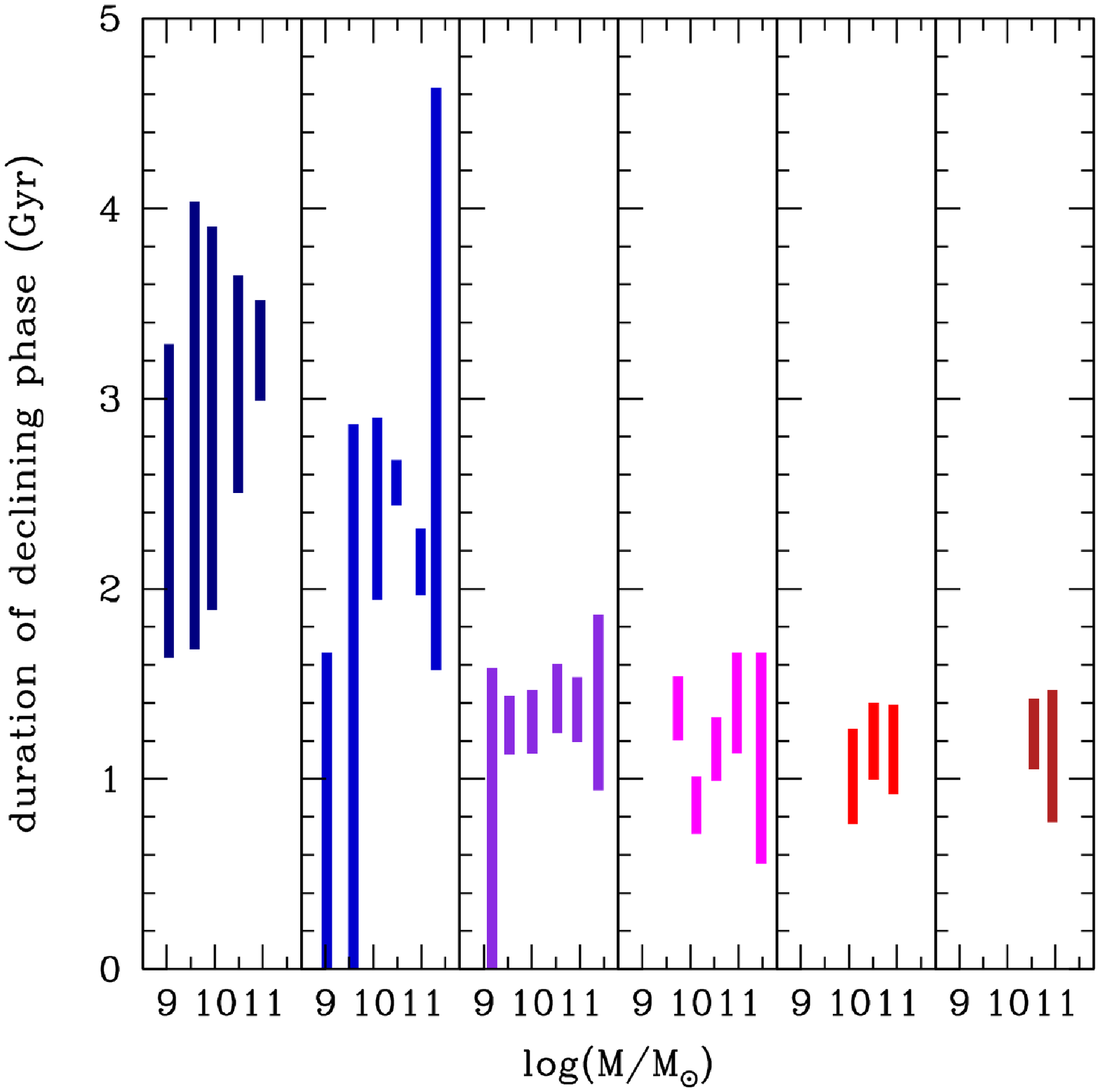}
\caption{The duration of the rising and declining phases of the median SFHs in Figure~\ref{fig:sfh} varies as a function of stellar mass and redshift. For each bin in stellar mass and redshift in Figure~\ref{fig:sfh}, we show, on the left-hand side, the distribution of the duration of the rising phase, and, on the right-hand side, the distribution of the duration of the declining phase, as a function of stellar mass. Each small panel represents a different redshift bin. Bars are color coded according to redshift.}
\label{fig:phases}
\end{center}
\end{figure*}

We measure the location of the peaks of the median SFHs in time in Figure~\ref{fig:sfh}. In the same way as for the widths, we bootstrap 100 times the galaxies in each bin and we measure the distribution of the lookback time at which the SFR is the highest. In Figure~\ref{fig:peak}, we plot the 25-to-75 percentiles of the distribution of the lookback time of the peak of the median SFHs as a function of stellar mass and redshift.

At low redshift, the lookback time of the peak is a function of stellar mass: low-mass galaxies reach their peak later in their lifetimes (around 5--6 Gyr in lookback time) than high-mass galaxies, which reach their peak at 8--9 Gyr in lookback time. At high redshifts, this trend appears to flatten. Because of the completeness limits of the sample, we do not have much information for low masses at high redshift. The trend between the lookback time of the peak and stellar mass is clearly visible also when the lookback time of the peak is expressed in units of the age of the Universe.

\subsection{Duration of the rising and declining phases of the SFHs}
\label{sec:risedec}

We measure the duration of the rising and the declining phases of the median SFHs of Figure~\ref{fig:sfh}. We define the duration of the rising phase to be the time between when the median SFH forms 10\% of its total stellar mass and when it reaches its peak. We define the declining phase to be the time between the peak and when 90\% of the total stellar mass is formed. In the same way as for the measurements of the width and the location of the peak of the median SFHs, we use a bootstrapping procedure to measure the duration of the rising and declining phases.

Figure~\ref{fig:phases} shows the duration of the rising and declining phases as a function of stellar mass and redshift. In low-redshift galaxies, the star formation can increase for a long period of time (3--4 Gyr) before moving to quiescence. High-mass galaxies instead seem to grow and reach their peak of star formation faster in about 1 Gyr. The declining phase is then fast in low-mass galaxies ($\sim 1$ Gyr) and slow in high-mass galaxies ($\sim 3$ Gyr), at low redshift. At high redshifts, the Universe is young and thus both phases have to happen rapidly by construction ($\sim 1$ Gyr). We note that, at high-redshift, galaxies might also be shutting off the star formation slowly. Such galaxies would not have time to become completely quiescent by the time of observation and thus are excluded from our sample.

\section{Discussion}
\label{sec:discussion}

\begin{figure*}
\begin{center}
\includegraphics[width=\textwidth]{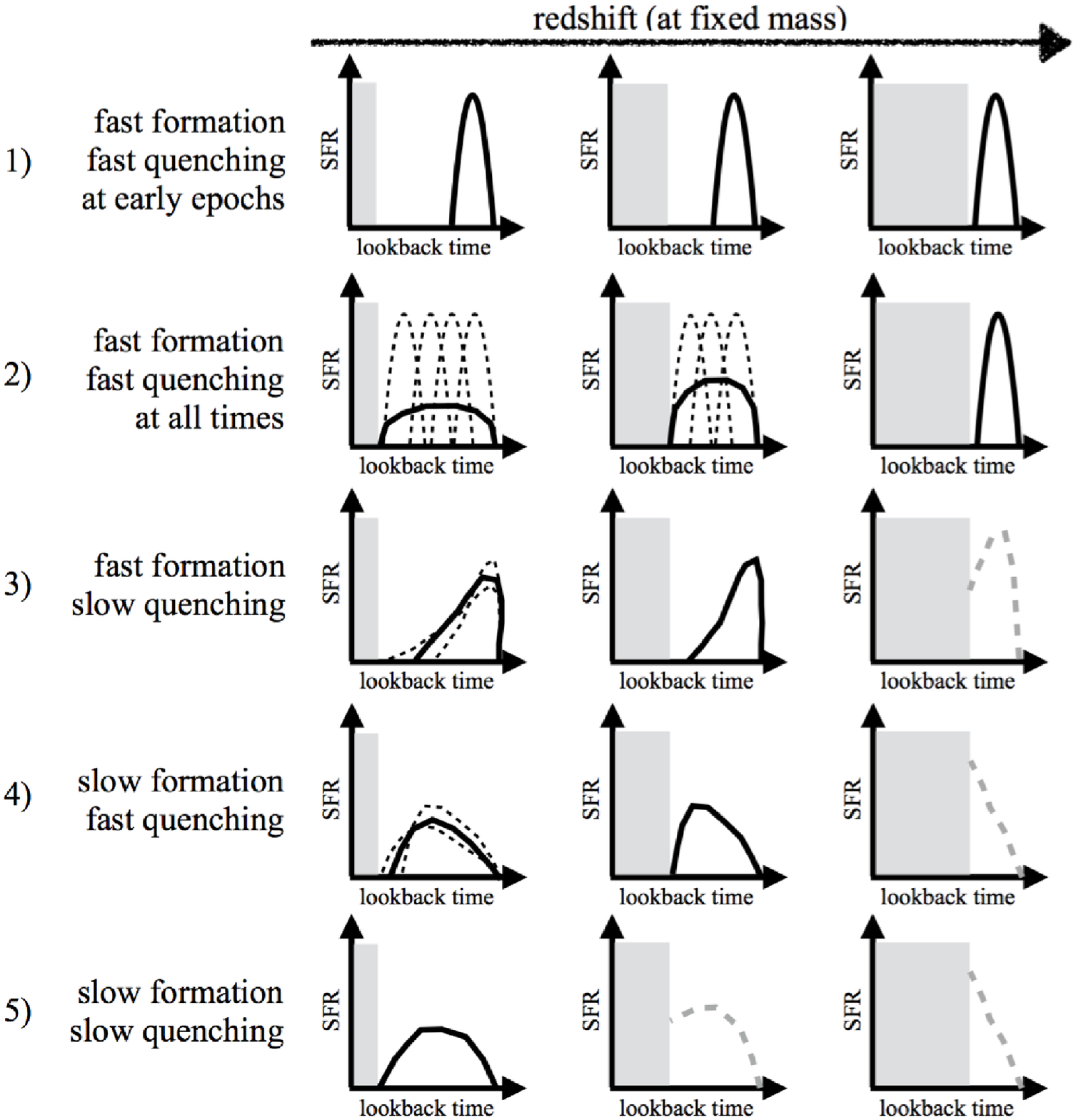}
\caption{Five possible scenarios to explain the variety of shapes of quiescent galaxy SFHs are shown. Each row corresponds to a different scenario and the format mimics that of Figure~\ref{fig:sfh}, i.e., redshift grows from left to right at fixed stellar mass and the portion of the Universe that is not observable at a particular redshift is shaded grey. In the same way as in Figure~\ref{fig:sfh}, the observed median SFHs (black thick lines) are combinations of individual SFHs (black dashed lines). The scenarios are: 1) Galaxies form fast and quench fast at early times. 2) Galaxies form fast and quench fast at different epochs. For this the observed median SFH appears constant with fast rising and declining phases. 3) Galaxies form fast and quench slowly. These galaxies reach their peak early and have a long tale of decreasing star formation. 4) Galaxies increase their star formation for long periods of time, reach their peak late in their life and become quiescent fast. 5) Galaxies form and quench slowly. The high-redshift Universe is not old enough to allow galaxies to reach quiescence when at least one of the two phases happens slowly, thus galaxies would only be star forming at observation (grey dashed lines).}
\label{fig:cartoon}
\end{center}
\end{figure*}

The way galaxies evolve from the blue cloud to the red sequence is most likely an interplay of different mechanisms, e.g. dry and wet mergers, AGN and stellar feedback, catastrophic events, residual-gas consumption (see \citealt{somerville2015} for a review). Each of these mechanisms cause changes in the way galaxies form (and stop forming) stars. By measuring the SFHs of galaxies, we can quantify what are the dominant modes of quenching and on what timescales they happen. Given the trends shown in Section~\ref{sec:results} and Figures~\ref{fig:width}, \ref{fig:peak}, and \ref{fig:phases}, we now discuss what they tell us about quenching mechanisms. We first describe possible scenarios for the evolution of quiescent galaxies and then we associate them to the observations and to physical mechanisms. All our conclusions will apply to populations of galaxies, because we will compare the possible scenarios with the median SFHs we have derived. We remind the reader that single galaxies can evolve differently and their individual SFHs are likely more complex.

\subsection{A list of possible scenarios}

We show in Figure~\ref{fig:cartoon} a simple cartoon describing five possible scenarios that can explain the median SFHs we observe in Figure~\ref{fig:sfh}. We note that we refer to durations as absolute quantities here, but we will consider the evolution of the dynamical timescale when we connect the scenarios to the observations in Section~\ref{sec:interp}. The scenarios are as follows.

\begin{enumerate}
\item \textit{Fast formation and fast quenching at early epochs}. In this picture galaxies form fast and quench fast in the first billion years of the Universe similar to the way local early-type galaxies were interpreted a few decades ago (\citealt{baade1963}).

\item \textit{Fast formation and fast quenching at all times}. In this picture, galaxies form fast and quench fast at different times. The distribution in time of these SFHs can take on a number of different shapes. For example, a combination of these spiky SFHs in time can result in median SFHs that are flat, as depicted in Figure~\ref{fig:cartoon}. 

\item \textit{Fast formation and slow quenching}. In this picture, galaxies undergo a rapid formation and a slow decline in SFR. A distribution of different quenching timescales would produce a large scatter in the declining phase of the median SFHs.

\item \textit{Slow formation and fast quenching}. In this picture, the SFR grows slowly for long periods of time, the peak of star formation is reached relatively late, and the quenching phase happens fast.

\item \textit{Slow formation and slow quenching}. In this picture, both formation and quenching are slow processes.
\end{enumerate}

\subsection{Constraining quenching scenarios with observations}
\label{sec:interp}

\subsubsection{Confirming previous results}

If we compare the median SFHs in Figure~\ref{fig:sfh} with the scenarios presented in Figure~\ref{fig:cartoon}, we can argue that a single scenario cannot explain all the different behaviors. We therefore infer that the red sequence most likely builds through a combination of different mechanisms, as pointed out already in previous works, e.g. \citet{willmer2006}, \citet{bell2006}, \citet{faber2007}, \citet{skelton2012}. Furthermore, we find downsizing in the evolution of quiescent galaxies, i.e. massive galaxies evolve faster and reach their peak of star formation earlier than low-mass galaxies. This has also been seen previously, e.g., \citet{thomas2005}.

\subsubsection{New findings}

We find that, on average, the observed populations of quiescent galaxies take as much time as they are allowed by the age of the Universe to quench, i.e. the median SFHs always extend along the full age of the Universe at the redshift of observation. The large confidence ranges reflect the intrinsic variety of star formation histories in each bin (see also \citealt{dressler2016}). Interestingly, we observe that the widths of the SFHs are a constant fraction of the age of the Universe at all redshifts (Figure~\ref{fig:width}, right-hand side). This could be a consequence of the fact that the SFHs extend over long periods of time, and they look quasi bell-shaped, thus the FWHM has to be somewhere around 0.5 times the age of the Universe. However, it could also be fundamental. For example, using cosmological simulations, \citet{tacchella2016} found that star-forming galaxies oscillate about the star-formation main sequence  on timescales $\sim 0.4--0.5 \, t_{Hubble}$. Galaxies form the bulk of their stellar mass when on the main sequence, thus this prediction would agree with what we find from observations. In a following paper, we will compare these measurements with different cosmological simulations and assess whether this result is indeed fundamental. Using a different semi-analytical model, \cite{tonini2012} find a similar result when comparing color predictions of bright cluster galaxies to different observed samples at $z<1.6$.

At high redshift (Figure~\ref{fig:sfh}, last columns on the right), galaxies have to form and quench fast (similar to scenario 1) in order to be quiescent due to the short age of the Universe at these redshifts. Given that the dynamical time of galaxies evolves with redshift, timescales of about 1 Gyr at $z\sim2$ are consistent with an evolving halo quenching mass, i.e., a critical mass above which gas accretion from cold flows is prevented (\citealt{keres2005}, \citealt{dekel2006}, \citealt{mitra2015}). At this redshift, any galaxies, that are in the process of shutting off their star formation on slow timescales would not be quiescent and therefore would not be in our observational sample.

At low-redshift (Figure~\ref{fig:sfh}, first columns on the left), the median SFHs of quiescent galaxies extend for long periods of time and thus are not fully consistent with scenario 1, which implies passive evolution of high-redshift quenched galaxies (although a few individual galaxies could do this). This finding is in agreement with previous works, such as \citet{worthey1992}, \citet{faber1995}, \citet{trager2000}, \citet{harker2006}, \citet{schiavon2006}, \citet{ruhland2009}, and \citet{gallazzi2014}. Assuming that all quiescent galaxies we observe at $z\sim2$ stay quiescent to $z\sim0.2$, from number density calculations (i.e., calculating how many of the galaxies we observe in the big volume at high redshift are observable in the smaller volume at low redshift), we find that at low redshift the contribution of galaxies already quenched at $z\sim 2$ is only 13\%. This 13\% is an upper limit since the high-redshift quiescent galaxies can in principle undergo merging and/or gas accretion and re-ignite star formation. Such a percentage is not large and thus we expect not to have a large contribution from them in the median SFHs of low-redshift quiescent galaxies. 

Low-redshift, low-mass galaxies (Figure~\ref{fig:sfh}, panels in the top left) on average seem to undergo a period of constant star formation in between the rising and declining phases (similar to scenario 2). We could also argue that these galaxies take a long time to form, fed by continuous gas accretion (as predicted by \citealt{behroozi2013} and \citealt{moster2013}), which is more similar to scenario 4. Given that the star formation is almost constant for a long period of time and because we are selecting quiescent galaxies, the declining phase must be fast, thus probably catastrophic. By catastrophic events, we mean processes such as strong feedback (\citealt{croton2006}, \citealt{somerville2008}) and environmental effects (\citealt{peng2010}). Such low-mass galaxies might also be satellites of big halos and therefore subject to gas stripping that shuts off their star formation quickly (e.g., \citealt{wetzel2013}, \citealt{woo2013}). The SFHs of individual low-mass galaxies could also be very bursty \citep{kauffmann2014}. A galaxy with a bursty SFH could experience long quiescence periods if the accretion rate slows down. The average SFH of these type of galaxies would show a period of fairly constant star formation (the single bursts are washed out in the average) and a fast decline, and would thus be consistent with what we observe at low stellar masses.

Finally, low-redshift, high-mass galaxies (Figure~\ref{fig:sfh}, panels in the bottom left) show, on average, a fast formation phase and a slow quenching phase, ruling out scenarios 4 and 5. The existence of mass quenching or another process which makes SFR become inefficient could explain such SFHs. If gas accretion is prevented, galaxies would reach quiescence by consuming their reservoir of gas. This process is consistent with a slow quenching phase and would also explain the turnover of the galaxy star-formation main sequence at high masses and low redshifts (\citealt{whitaker2014}, \citealt{gavazzi2015}, \citealt{schreiber2015}).

A combination of galaxies that form and quench fast (scenario 2) at different times can reproduce slow and fast rising and quenching phases. For example, if the number of galaxies that form and quench fast is large at high-redshift and progressively decreases in time, the median SFHs would look like the one in the scenario 3 (fast rising and slow quenching). However, a fast quenching mechanism at all times and masses is not consistent with the turnover of the star-formation main sequence, because galaxies would ``jump'' from the blue to the red sequence and we would hardly observe any in the intermediate phase between being star-forming and quenched. Furthermore, we observe individual galaxies with long declining phases when the age of the Universe is sufficiently long, therefore we rule out scenario 2 for the majority of the galaxies at low redshift.

\section{Conclusions}
\label{sec:concl}

Although there has been much progress in understanding how galaxies evolve to the present day, we still do not understand how and when galaxies stop forming stars and become quiescent. Using the CANDELS dataset, we measure the SFHs of 845 quiescent galaxies at $0.2<z<2.1$. Our galaxy SED models allow us to extract full SFHs of galaxies from the fossil record embedded in their photometric measurements. Specifically, our model library of SEDs adopts complex SFHs and metal enrichment histories from the semi-analytic post-processing of a large cosmological simulation. These histories are consistent with the observed evolution in the mass-metallicity relation. Moreover, adoption of such histories acts as an important prior, limiting the effects of degeneracies in age and metallicity.

Using the individual SFHs we measure from the observations, we compute median SFHs of quiescent galaxies in bins of stellar mass and redshift. They are quasi bell-shaped: SFR rises, reaches a peak, and then declines towards quiescence.

We find that on average galaxies take as much time as they can to quench. At high redshift, when the Universe is young, the median SFHs can only be narrow and the quenching phase fast. Given that the dynamical time of galaxies evolve with redshift, timescales of about 1 Gyr at $z\sim2$ are consistent with an evolving halo quenching mass and do not necessarily require catastrophic events such as strong feedback and violent environmental effects.

Low-redshift, low-mass galaxies, on average, reach their peak of star formation later in their lifetimes compared with high-mass galaxies. This is downsizing. Given this and the fact that we are selecting quiescent galaxies, low-mass galaxies, on average, shut off their star formation fast. A catastrophic event (such as strong feedback) might be required to reproduce a fast declining phase at low redshift. Such low-mass galaxies might also be satellites in big halos subject to gas stripping.

Low-redshift, high-mass galaxies show on average long declining phases. Galaxies could reach a critical mass which prevents new cold gas to inflow. Such long declining phases would then be explained by slow consumption of residual gas after the inflow of new gas stops. The residual gas is consumed slowly and the star formation decreases gradually. This is consistent with the turn over of the star-formation main sequence at the high-mass end, as observed in previous works.

In this work, we have uncovered the dependences of the SFHs on stellar mass and redshift. We will explore the characteristics of individual SFHs in relation to environment and morphology, in a future work. The possible correlations between SFH shape and environment is of particular interest, although larger fields would be required. Also, we will examine the constraints on the SFHs of both star-forming and quiescent galaxies. By studying the entire population, we will be able to derive the probability that a galaxy shuts off its star formation and moves to the quiescent population as a function of stellar mass and redshift. The comparison of our constraints with detailed simulations of galaxy formation and evolution will be crucial to find the realistic scenarios that best resembles the observations, and thus quantify the importance of the different evolutionary mechanisms.

\acknowledgments

We thank the Referee for the nice and helpful report.
We thank Stephane Charlot, Julianne Dalcanton, and Steven Willner for useful discussions.
CP acknowledges support by an appointment to the NASA Postdoctoral Program at the Goddard Space Flight Center, administered by USRA through a contract with NASA.
CP is grateful for the support by SAK through HST grant AR-12828.001 and a Director's Discretionary Research Fund (DDRF).
This work is based on observations taken by the CANDELS Multi-Cycle Treasury Program with the NASA/ESA HST, which is operated by the Association of Universities for Research in Astronomy, Inc., under NASA contract NAS5-26555.

\def\aj{AJ}
\def\araa{ARA\&A}
\def\apj{ApJ}
\def\apjl{ApJ}
\def\apjs{ApJS}
\def\apss{Ap\&SS}
\def\aap{A\&A}
\def\aapr{A\&A~Rev.}
\def\aaps{A\&AS}
\def\mnras{MNRAS}
\def\pasp{PASP}
\def\pasj{PASJ}
\def\qjras{QJRAS}
\def\nat{Nature}

\def\aplett{Astrophys.~Lett.}
\def\aas{AAS}
\let\astap=\aap
\let\apjlett=\apjl
\let\apjsupp=\apjs
\let\applopt=\ao

~\\

\section*{Appendix A: The star formation histories of quiescent galaxies in functional forms}
\label{app:dtau}
\renewcommand\thefigure{A\arabic{figure}}
\setcounter{figure}{0}
\renewcommand\thetable{A\arabic{table}}
\setcounter{table}{0}

The median SFHs we measure in Figure~\ref{fig:sfh} are fairly smooth and thus they could potentially be represented by analytic functions. The most common functions in the literature are exponentially declining functions of time (commonly called `tau-models', $\exp(-t/\tau)$). These functions can provide an acceptable approximation when the peak of star formation happens at very early times. We observe however that the SFR rises for a significant amount of time after formation, in all bins of stellar mass and redshift. We thus cannot properly represent the median SFHs we observe with simple exponentially declining tau-models.

Another analytic function sometimes used to model the SFHs of galaxies consist of a combination of a linear rising function and an exponentially declining function (commonly called `delayed tau-model', $t\,\exp(-t/\tau)$). This parametrization allows for a rising function, thus is more appropriate to reproduce the median SFHs we observe. However, this function does not allow for enough freedom in the position of the peak of star formation. If we take the derivative of the function and we set it to zero, we find that the peak happens at $t=\tau$. This means that the SFHs can either peak early and show a fast declining phase, or they can peak late and show a slow declining phase. A late peak and a slow declining phase would not produce a quiescent galaxy and the only feasible option would be a delayed tau-model with a short $\tau$. Most of the median SFHs we observe do not show such short declining phases and the peaks of star formation do not happen until at least 1 Gyr. The simple combination of linear and exponential decline is thus still not enough to reproduce what we observe.

We need to introduce at least one more parameter to allow for enough freedom in both the position of the peak of star formation and the duration of the declining phase. To this aim, we choose two analytic functions that reasonably reproduce the shape of most of the median SFHs in Figure~\ref{fig:sfh}: a combination of a power law and an exponentially declining function,
$$\psi(t)\propto t^d\exp(-t/\tau),$$
and a double power law
$$\psi(t)\propto [(t/\tau)^B + (t/\tau)^{-C}]^{-1}$$
as suggested by \citet{behroozi2013}. We show in Figure~\ref{fig:sfhfit} the fits with the two analytic functions to the observed median SFHs. The values of the best-fit parameters are reported in Table~\ref{tab:dtau}.

Figure~\ref{fig:sfhfit} shows that both analytic functions are reasonable approximations of the median SFHs we measure from the data. When the peak of star formation is not well defined and there is a period of roughly constant star formation, these functions fail to reproduce the shape. This happens especially at low stellar masses and low redshifts. We stress that these fits are valid only for median SFHs of quiescent galaxies. Individual galaxies can form and evolve with different timescales and most of all they can be characterized by episodes of star formation (e.g., rapid bursts) that cannot be captured by simple analytic functions.

{\renewcommand{\arraystretch}{1.2}
\begin{table*}
\begin{center}
\caption{Best-fit parameters of the two analytic functions: $d$ and $\tau$ for the power law and exponentially declining function (first line of every bin); $B$, $C$, and $\tau$ for the double power law (second line of every bin).}
\label{tab:dtau}
\begin{tabular}{l | c c c c c c}
\hline
\hline
Stellar mass	& \multicolumn{6}{c}{Redshift}\\
$\log(M_{\ast}/M_{\sun})$	& $0.2<z<0.5$ 	& $0.5<z<0.7$ 	& $0.7<z<1.0$	& $1.0<z<1.3$	& $1.3<z<1.7$	& $1.7<z<2.1$	\\
\hline
$\sim 9.0 $ &	7.6; 0.5		& 3.8; 1.2		& 2.6; 1.3		& ---				& ---			& ---	\\
		&	18.6; 0.8; 6.7	& 3.2; 1.8; 5.2	& 4.0; 1.4; 4.2	& ---				& ---			& --- \\
$\sim 9.5 $ & 	6.0; 0.7		& 2.6; 1.1		& 7.8; 0.4		& ---				& ---			& ---	\\
		&	19.8; 1.0; 8.1	& 6.8; 0.6; 5.3	& 12.0; 1.4; 4.8	& ---				& ---			& --- \\
$\sim 10.0 $ & 	8.2; 0.5		& 8.4; 0.4		& 9.8; 0.3		& 5.6; 0.4			& 3.4; 0.5		& ---	\\
		&	14.0; 0.8; 6.9	& 12.2; 1.2; 5.5	& 16.2; 1.0; 4.8	& 13.2; 1.2; 3.8		& 6.6; 1.0; 2.8	& --- \\
$\sim 10.5 $ & 	7.2; 0.6		& 5.8; 0.5		& 6.0	; 0.5		& 3.6; 0.6			& 2.6; 0.7		& 2.2; 0.6		\\
		&	11.2; 1.2; 7.1	& 10.4; 1.2; 4.9	& 10.4; 1.2; 4.8	& 9.6; 1.0; 2.7		& 5.0; 1.2; 2.7	& 3.8; 1.0; 2.0	\\
$\sim 11.0 $ & 	5.0; 0.7		& 7.4; 0.4		& 7.4; 0.4		& 3.6; 0.7			& 2.8; 0.7		& 2.4; 0.6	\\
		&	7.4; 1.0; 5.6	& 12.6; 0.8; 5.2	& 11.6; 1.2; 4.7	& 5.8; 1.4; 3.5		& 2.4; 1.6; 2.2	& 3.4; 1.0; 2.1	\\
$\sim 11.5 $ & ---			& 1.4; 1.3		& 7.0; 0.4		& 5.0; 0.4			& ---			& ---	\\
		&	---			& 2.8; 0.8; 3.0	& 12.6; 1.2; 4.6	& 10.6; 0.8; 3.5		& ---			& --- \\
\hline
\end{tabular}
\end{center}
\end{table*}

\begin{figure*}
\begin{center}
\includegraphics[width=0.9\textwidth]{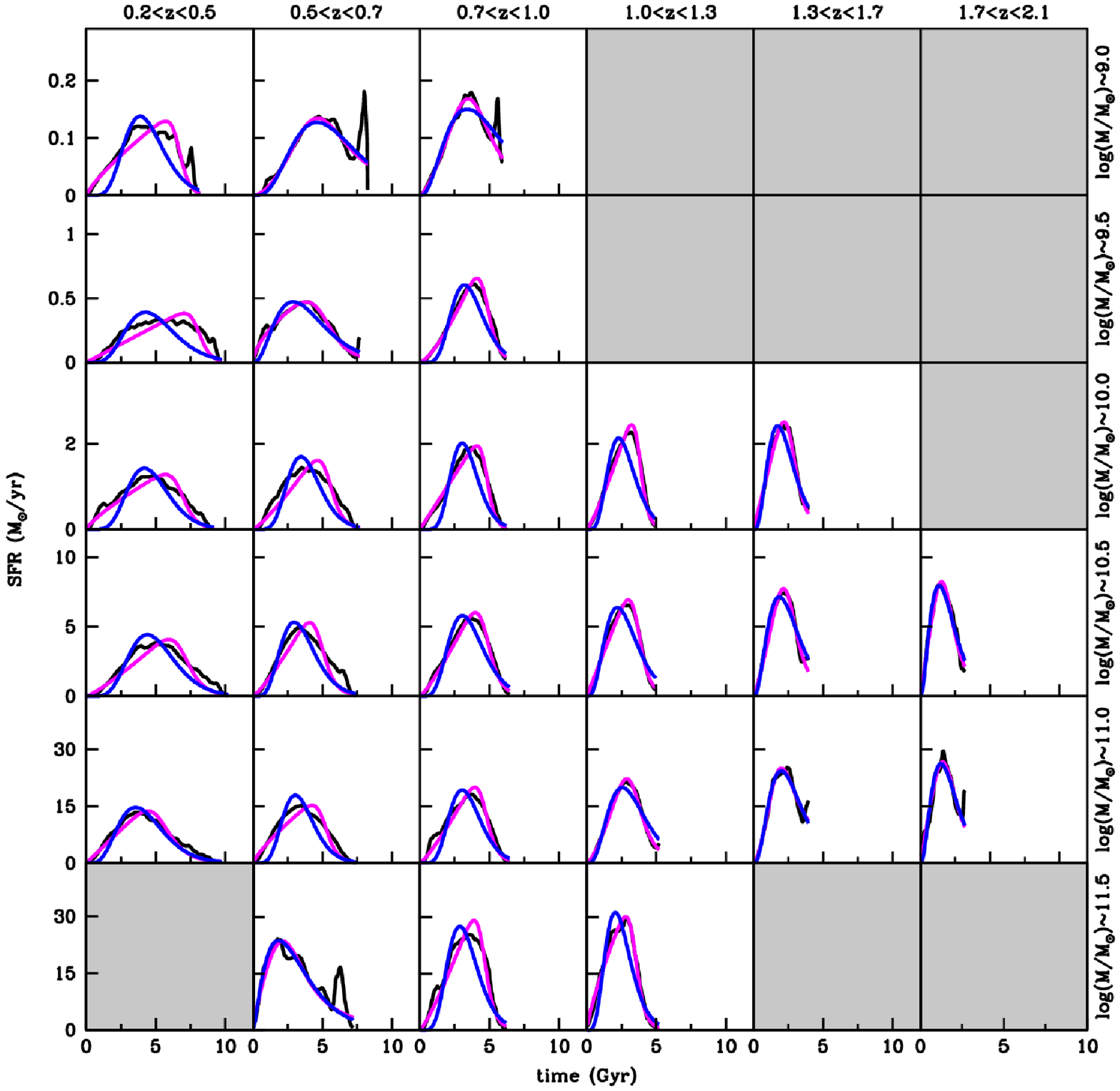}
\caption{Fits with analytic functions to the median SFHs of Figure~\ref{fig:sfh} (black). Time traces the age of the Universe here, thus it runs in the opposite direction of the lookback time. In blue, a combination of a power law and an exponentially declining function. In magenta, a double power law.}
\label{fig:sfhfit}
\end{center}
\end{figure*}
}

\end{document}